\documentclass[useAMS,usenatbib]{mn2e}
\usepackage{url,times,graphicx,amsmath,amsfonts,amssymb,aas_macros,color,epsfig,epstopdf}
\usepackage{xcolor}

\def \etal {et~al.~}

\newcommand{\hMpc}{{\ifmmode{h^{-1}{\rm Mpc}}\else{$h^{-1}$Mpc}\fi}}
\newcommand{\hkpc}{{\ifmmode{h^{-1}{\rm kpc}}\else{$h^{-1}$kpc}\fi}}
\newcommand{\kpc}{{\ifmmode{ {\rm kpc} }\else{{\rm kpc}}\fi}}
\newcommand{\kms}{{\ifmmode{ {\rm km\,s^{-1}} }\else{ ${\rm km\,s^{-1}}$ }\fi}}
\newcommand{\hMsun}{{\ifmmode{h^{-1}{\rm {M_{\odot}}}}\else{$h^{-1}{\rm{M_{\odot}}}$}\fi}}
\newcommand{\Msun}{{\ifmmode{{\rm M}_{\odot}}\else{${\rm M}_{\odot}$}\fi}}
\newcommand{\Mhalo}{{\ifmmode{M_{\rm halo}}\else{$M_{\rm halo}$}\fi}}
\newcommand{\Rvir}{{\ifmmode{R_{\rm vir}}\else{$R_{\rm vir}$}\fi}}
\newcommand{\Mstar}{{\ifmmode{M_{\star}}\else{$M_{\star}$}\fi}}
\newcommand{\Vrot}{{\ifmmode{V_{\rm rot}}\else{$V_{\rm rot}$}\fi}}
\newcommand{\ltsima}{$\; \buildrel < \over \sim \;$}
\newcommand{\gtsima}{$\; \buildrel > \over \sim \;$}
\newcommand{\lsim}{\lower.5ex\hbox{\ltsima}}
\newcommand{\gsim}{\lower.5ex\hbox{\gtsima}}

\def\lesssim{\mathrel{\hbox{\rlap{\hbox{\lower4pt\hbox{$\sim$}}}\hbox{$<$}}}}
\def\gtrsim{\mathrel{\hbox{\rlap{\hbox{\lower4pt\hbox{$\sim$}}}\hbox{$>$}}}}

\newcommand{\beq}{\begin{equation}}
\newcommand{\eeq}{\end{equation}}
\def\beqa{\begin{eqnarray}}
\def\eeqa{\end{eqnarray}}
\def\LCDM{\ensuremath{\Lambda}CDM}

\def\head{ \vbox to 0pt{\vss \hbox to 0pt{\hskip 440pt\rm
      LA-UR-10-07069\hss} \vskip 25pt}}
\def \xoff {\ifmmode x_{\rm off} \else $x_{\rm off}$ \fi}
\def \rhorms {\ifmmode \rho_{\rm rms} \else $\rho_{\rm rms}$ \fi}

\def \kms {\ifmmode  \,\rm km\,s^{-1} \else $\,\rm km\,s^{-1}  $ \fi }
\def \kpc {\ifmmode  {\rm kpc}  \else ${\rm  kpc}$ \fi  }  
\def \hkpc {\ifmmode  {h^{-1}\rm kpc}  \else ${h^{-1}\rm kpc}$ \fi  }  
\def \hMpc {\ifmmode  {h^{-1}\rm Mpc}  \else ${h^{-1}\rm Mpc}$ \fi  }  
\def \Mpch {\ifmmode  {h^{-1}\rm Mpc}  \else ${h^{-1}\rm Mpc}$ \fi  }  
\def \Msun {\ifmmode {\rm M}_{\odot} \else ${\rm M}_{\odot}$ \fi} 
\def \hMsun {\ifmmode h^{-1}\,\rm M_{\odot} \else $h^{-1}\,\rm M_{\odot}$ \fi}

\def \LCDM {\ifmmode \Lambda{\rm CDM} \else $\Lambda{\rm CDM}$ \fi}
\def \sig8 {\ifmmode \sigma_8 \else $\sigma_8$ \fi} 
\def \OmegaM {\ifmmode \Omega_{\rm m} \else $\Omega_{\rm m}$ \fi} 
\def \Omegab {\ifmmode \Omega_{\rm b} \else $\Omega_{\rm b}$ \fi} 
\def \OmegaL {\ifmmode \Omega_{\rm \Lambda} \else $\Omega_{\rm \Lambda}$\fi} 
\def \Deltavir {\ifmmode \Delta_{\rm vir} \else $\Delta_{\rm vir}$ \fi}
\def \rhocrit {\ifmmode \rho_{\rm crit} \else $\rho_{\rm crit}$ \fi}
\def \rhou {\ifmmode \rho_{\rm u} \else $\rho_{\rm u}$ \fi}
\def \zc {\ifmmode z_{\rm c} \else $z_{\rm c}$ \fi}

\def\head{
.ps \vbox to 0pt{\vss
                   \hbox to 0pt{\hskip 440pt\rm LA-UR-10-07069\hss}
                  \vskip 25pt}}
\title[Strongly Coupled Cosmologies I]
{Strongly Coupled Dark Energy Cosmologies:\\ preserving ${\bf \Lambda}$CDM
  success and easing low scale problems \\
I -- Linear theory revisited}

\author[S.A. Bonometto \etal] { Silvio A.
  Bonometto$^{1,2}$\thanks{E-mail: bonometto@oats.inaf.it}, Roberto
  Mainini$^3$, Andrea V. Macci\`o$^4$\\
$^1$Physics Department, Astronomy Unit, Trieste University, Via
  Tiepolo 11, I-34143 Trieste, Italy \\ $^2$ I.N.A.F., Osservatorio
  Astronomico di Trieste, I-34143 Trieste, Italy \\ 
  $^3$ Physics
  Department G.~Occhialini, Milano--Bicocca University, Piazza della Scienza 3, I-20126
  Milano, Italy\\ $^4$Max-Planck-Institut f\"ur Astronomie,
  K\"onigstuhl 17, D-69117 Heidelberg, Germany}

\begin{document}

\date{Accepted XXXX . Received XXXX; in original form XXXX}

\pagerange{\pageref{firstpage}--\pageref{lastpage}} \pubyear{2010}

\maketitle

\label{firstpage}


\begin{abstract}

In this first paper we discuss the linear theory and the background
evolution of a new class of models we dub SCDEW: Strongly
  Coupled DE, plus WDM.  In these models, WDM dominates today's  
matter density; like baryons, WDM is uncoupled. Dark Energy is a
scalar field $\Phi$; its coupling to ancillary CDM, whose today's
  density is $\ll 1\, \%$, is an essential model feature.
Such coupling, in fact, allows the formation of cosmic structures, in
spite of very low WDM particle masses ($\sim 100$ eV).
SCDEW models yields Cosmic Microwave Background and linear Large Scale
features substantially undistinguishable from $\Lambda$CDM,
but thanks to the very low WDM masses  they 
strongly alleviate \LCDM issues on small scales, as confirmed via
numerical simulations in the II associated paper.
Moreover SCDEW cosmologies significantly ease the coincidence and fine
tuning problems of $\Lambda$CDM and, by using a field theory approach,
we also outline possible links with inflationary models.
We also discuss a possible fading of the coupling at low redshifts
which prevents non linearities on the CDM component to cause
  computational problems. The (possible) low--$z$ coupling
  suppression, its mechanism, and its consequences are however still
  open questions --not necessarily problems-- for SCDEW models.
  The coupling intensity and the WDM particle mass, although
  being extra parameters in respect to \LCDM, are found to be
  substantially constrained {\it a priori} so that, if SCDEW is the
  underlying cosmology, we expect most data to fit also \LCDM
  predictions.

\end{abstract}

\noindent
\begin{keywords}
large--scale structure of Universe, dark matter, dark energy - Galaxy: evolution, formation 
 \end{keywords}

\section{Introduction} \label{sec:introduction}

When SNIa Hubble diagrams revitalized $\Lambda$CDM, one could hardly
guess how successful this cosmology would be in meeting cosmological
data. In spite of its coincidence and fine tuning problems, therefore,
$\Lambda$CDM is surely the benchmark for any attempt to improve our
understanding of cosmology.  

In this series of works we present a detailed analysis of a new class
of cosmologies that preserves all successes of $\Lambda$CDM models on
large and intermediate scales. In turn, they improve the agreement
with the observed dark matter distribution on small scales
\citep[e.g.][]{Oh2011,walker}, at the same time easing the coincidence
and fine tuning problems of $\Lambda$CDM, while possibly launching a
bridge between the reheating stages in the late inflationary regime
and today's Dark Energy (DE) nature.

These cosmological models are based on the presence of three dark
components: i) a very--warm dark matter component (WDM), uncoupled (like
the baryons), and constituting the observable dark matter at the
present time; ii) a scalar field $\Phi$ which acts as Dark Energy;
iii) a peculiar Cold Dark Matter component (CDM), whose coupling
to $\Phi$ played an essential role in cosmic history, but whose
density at $z=0$ naturally became almost negligible. We dubbed these
models SCDEW (Strongly Coupled DE + WDM).

The basic features of these models were discussed in two previous
papers (\citet{bonometto2012}; \citet{bonometto2014}; BSLV12 and BM14
respectively, hereafter). BSLV12 dealt with background components:
they showed that a purely kinetic scalar field $\Phi$ strongly coupled
to a CDM component would both exhibit densities $\propto a^{-4}$ ($a:$
scale factor), during the radiative eras, expanding along a tracker
solution, with (primeval) density parameters $\cal O$$(0.1\, \%)$. The
derelativisation of a further WDM component eventually turns these
components into quintessential DE and a tiny CDM component. The
radiative eras of these models are therefore characterized by three
low density components, in top of the {\it usual} $\gamma$'s and
$\nu$'s: WDM, coupled CDM and scalar field, accounting for constant
fractions of the overall density, and sharing similar densities. BM14
then deals with fluctuation evolution, showing that CMB anisotropy and
polarization spectra, in these models, are hardly distinguishable from
the $\Lambda$CDM benchmark, while coupled CDM fluctuations continue to
grow also between their entry in the horizon and matter--radiation
equality, being so able, in spite of the low CDM density, to
revitalize WDM fluctuations on scales suffering an early free
streaming, even when the WDM particle mass $m_{\rm w}$ is quite small.

The $\Phi$--CDM coupling is therefore an essential feature in the
early model evolution. After WDM has derelativized, instead, the
$\Phi$--CDM coupling is unessential, while leading to technical
difficulties, as $\delta_c$ (CDM fluctuation amplitudes) tend to
approach unity, developing then some early non-linearities, typically
involving $\ll 1\%$ of the total mass.  However, switching off the
$\Phi$--CDM coupling after the break of ``conformal invariance'',
yields more comfortably tractable models.

The plan of the paper is as follows. In the next Section we shall
formulate the Lagrangian theory for strongly coupled Dark Energy
cosmologies. It confirms the prediction on the coupled $\Phi$--CDM
component densities, while outlining possible links with inflationary
theories.  In Section 3 we discuss the evolution of background
parameters, focusing on the exit from the primeval stationary
regime. We also widen the range of models in respect to BSLV12 and
BM14 papers, by allowing for $\beta$ fading after the exit from
conformally invariant expansion. In Section 4 we discuss fluctuation
evolution, taking also into account a possible $\beta $ fading. The
effects of the early growth of CDM fluctuations are therefore
analised, both as cause for the restart of WDM and baryon
fluctuations, and for the possible formation of late CDM
structures. We also show that, in order to mimic $\Lambda$CDM
phenomenology through SCDEW models, the mass of WDM particles and the
coupling are significantly constrained. In Section 5 fluctuation
spectra are obtained and the model for N--body simulations of Paper II
is conveniently selected. The last Discussion Section outlines the
predictions of SCDEW models, namely for what concerns their
discrimination from $\Lambda$CDM, whose main results are however
faithfully reproduced.  We also outline how SCDEW cosmologies, besides
of easing low--scale $\Lambda$CDM conundrums, free us from its
coincidence paradox and fine tunings.

\section{Coupled Dark Energy in the early Universe} \label{sec:CDE}
    
There are direct evidences that Dark Matter (DM) is a physical cosmic
component, mostly clustering with observable baryons. Doubts have been
cast, on the contrary, on the true nature of Dark Energy (DE); here we
assume it to be a self--interacting scalar field $\Phi$, keeping
essentially unclustered.  In this scheme, both dark components
interact with baryons and radiations just gravitationally, so that the
energy {\it pseudo--}conservation equation
\begin{equation}
\label{conti0}
T^{(c)~\mu}_{~~~~\nu;\mu} + T^{(\Phi)~\mu}_{~~~\, ~~\nu;\mu} = 0
\end{equation}
holds (here $T^{(c,d)}_{\mu\nu}$ are the stress--energy tensors of DM
and DE, their traces reading $T^{(c,\Phi)}$). This sum can vanish
thanks to separate vanishings of both terms. The alternative option
that
\begin{equation}
T^{(\Phi)~\mu}_{~~~\, ~~\nu;\mu} = +C T^{(c)} \Phi_{,\nu}~,
~~~~~~~~~~
T^{(c)~\mu}_{~~~~\nu;\mu} =- C T^{(c)} \Phi_{,\nu}~,
\label{conti1}
\end{equation}
however, is also widely considered in the literature (see, e.g.,
\citet{ellis}; \citet{wetterich}; \citet{amendola1999}; \citet{amendola2002};
\citet{maccio2004}; \citet{baldi}),
together with other possible options for energy transfer between DM
and DE (see, e.g., \citet{lopez}). The coupling
\begin{equation}
 C = b/m_p= (16 \pi/3)^{1/2} \beta/m_p~,
\label{beta}
\end{equation}
sets the intensity of the energy flow from DM to DE. The option of
scale dependent $C$ is also discussed in the literature, namely in
connection with specific models (see, e.g., R.~Mainini \&
S.A.~Bonometto 2004). Let the background metric then read
\begin{equation}
ds^2 = a^2(\tau) (d\tau^2 - d\lambda^2)~,
\label{metric}
\end{equation}
($\tau:$ conformal time, $d\lambda:$ the spatial element);
eqs.~(\ref{conti1}) then yields
\begin{equation}
\ddot \Phi + 2{\dot a \over a} \dot \Phi = -a^2 V'+ C a^2 \rho_c
~,~~~~~~~ \dot \rho_c + 3 {\dot a \over a} \rho_c = -C \rho_c \dot \Phi~,
\label{eq6}
\end{equation}
$\rho_c$ being the DM density, while $V(\Phi)$ is a self--interaction
potential, as is required within quintessential DE models.

The rational is to allow for an energy flow from (cold)--DM to DE. In
this way, the field density could keep a steady fraction (some
permils) of CDM density during the whole cosmic expansion, in spite of
$\Phi$ being essentially kinetic above a suitable redshift $z_{\pm}$.
At $z_\pm$ the DE state parameter therefore shifts from $\sim -1$ to
$\sim +1$ and such shift is found to be a generic feature,
indipendently of the shape of $V(\Phi)$.  By keeping a significant
field density at high $z$, this option eases one of the coincidence
problems of $\Lambda$CDM.

\subsection{An early $\Phi$--CDM coupling}
Within a field theory context, a possible assumption is that CDM is a
non--relativistic Dirac spinor field $\psi$, interacting with $\Phi$
through a Yukawa--like lagrangian
\begin{equation}
{\cal L}_m = -\mu f(\Phi/m) \bar \psi \psi~;
\label{inter}
\end{equation}
here 2 mass scales, $m = m_p/b$ and $\mu = g\, m_p$ are introduced for
dimensional reasons, $m_p$ being the Planck mass. In particular, $b$
coincides with the coupling parameter in eq.~(\ref{beta}) (see below).

By assuming a kinetic part of the scalar field lagrangian $ {\cal L}_k
\sim \partial_\mu \Phi \, \partial_\mu \Phi~, $ its equation of motion
reads
\begin{equation}
\ddot \Phi + 2{\dot a \over a} \dot \Phi = -a^2 V' -
a^2 \rho_c {f' \over f}~,
\label{eqmot}
\end{equation}
once we work out that
\begin{equation}
\rho_c = -{\cal L}_m = \mu f(b\Phi/m_p) \bar \psi \psi~,
\end{equation}
in the absence of a significant kinetic term for the spinor quanta.
Notice that the number density operator for the spinor field
$n \propto \bar \psi \psi$, so that
\begin{equation}
\rho_c \propto f(\Phi/m) a^{-3}
\label{rhosc}
\end{equation}
according to the findings of \citet{das}.

Eq.~(\ref{eqmot}) is consistent with the former eq.~(\ref{eq6}) if
$f'/f = -b/m_p$ so that, if the function inserted in
the Lagrangian~(\ref{inter}) has the form
\begin{equation}
f = \exp(-b \Phi/m_p)
\label{fff}
\end{equation}
we re--obtain the coupled--DE theories of eqs.~(\ref{conti1}). Also
$\rho_c$, according to eq.~(\ref{rhosc}), then exhibits a sort of
exponential scaling, unless we assume that the argument of the
exponential
\begin{equation}
b \Phi/m_p = \ln(\tau/\tau_r)~,
\label{newansatz}
\end{equation}
$\tau_r$ being a (unconstrained) reference time; then $f =
\tau_r/\tau$ and
\begin{equation}
\dot \Phi = {m_p \over b \tau}~.
\label{dotf}
\end{equation}
Let us consider the radiative era, $\Phi$ bearing
essentially kinetic energy; then the r.h.s. of eq.~(\ref{eqmot}) reads
$a^2 \rho_c b/m_p$, while the whole equation means that $ \rho_c = (
m_p / a \tau )^2 $ (thence $\rho_c \propto a^{-4}~!$) or, equivalently,
\begin{equation}
\left( \dot a \over a \right)^2 = {8 \pi \over 3} {1 \over m_p^2}\,  a^2 
(2 \beta^2 \rho_c)~. 
\label{quasifried}
\end{equation}
By comparing this equation with  the Friedmann eq.~, we deduce
a constant early density parameter for CDM
\begin{equation}
\Omega_c \equiv {\rho_c \over \rho} = {1 \over 2 \beta^2}~.
\label{omegac}
\end{equation}
In turn, $\Phi$ being purely kinetic, owing to eq.~(\ref{dotf}), its
energy density
\begin{equation}
\rho_\Phi \equiv {\dot \Phi^2 \over 2 a^2} = 
 \left( m_p \over b \right)^2 {1 \over 2 a^2 \tau^2}
= {\rho_c \over 2}~,
\label{rhod}
\end{equation}
coinciding with its pressure. The state parameter of the future
quintessential field is then $w = +1$, while its {\it early}
density parameter
\begin{equation}
\Omega_\Phi = {\Omega_c \over 2} = {1 \over 4 \beta^2}~.
\end{equation}
In connection with the scale dependence found for $\rho_c$, let us
outline that, if we read $\cal L$$_m$ as a mass term of the Dirac
spinor, its mass scales $\propto a^{-1}$ (as though being {\it
  redshifted}). As a matter of fact, both $\Phi$ and CDM energy
densities scale as $a^{-4}$. In the absence of coupling they would
scale as $a^{-6}$ and $a^{-3}$, respectively. The flow of energy from
CDM to $\Phi$ is tuned to yield a faster (slower) CDM ($\Phi$ field)
dilution.

The key issue, however, shown in BSLV12, is that these conditions are
an attractor: even significantly perturbing ``initial conditions'',
the cosmic evolution rapidly settles on the behavior here described.
Accordingly, such kind of evolution could last since a very early
epoch, even since the late inflationary stages. 

We might tentatively assume that cosmic reheating was due to the very
$\Phi$ field decaying into $\psi$ field quanta, a process stopping as
soon as the attractor solution is attained. Interactions of $\psi$
with other fields would then allow the $\Phi$ field energy to reheat
other cosmic components, including both $\gamma$'s \& $\nu$'s, as well
as any other component then belonging to the {\it thermal soup}.

The ensuing self--similar expansion, that can be defined {\it
  conformally invariant} (in the sense discussed, e.g., by Parker
1969,1971), can only be broken if a component scaling differently from
$a^{-4}$ achieves a significant density. If the attractor solution is
followed since inflation, $\Phi$ has had just a logarithmic time
dependence since then. Its today's value, therefore, is just $\sim 60$
times its value at the end of the reheating stages.

\section{Exit from the stationary regime}\label{sec:Exit}

In any reasonable cosmological model, baryon density would eventually
break the conformally invariant expansion at a redshift $z_b \gtrsim
500$. Adding just a baryon component to radiative components (coupled
and uncoupled) is however insufficient to meet observations. Among
viable possibilities, BM14 outlined the option of including a Warm DM
(WDM) component of thermal origin, with a temperature parameter
$T_{\rm w}$.  Its derelativization occurs when $T_{\rm w} \sim m_{\rm
  w}$ (mass of WDM quanta) at a redshift $z_{\rm w}$. At $z \gg z_{\rm
  w}$, being $P_{\rm w} \simeq \rho_{\rm w}/3$, it is $\rho_{\rm w}
\propto T_{\rm w}^4$ and WDM is one of the components of the {\it
  thermal soup}; then, at $z \ll z_{\rm w}$, being $P_{\rm w} \ll
\rho_{\rm w}$ and therefore $\rho_{\rm w} \propto T_{\rm w}^3$, WDM
shall overcome the radiative component density so that the early
stationary regime reaches an end.

This also sets an end to the attractor solution of the coupled
components. During their successive evolution, as in any model
allowing for CDM--DE coupling with small or large $\beta$, the state
parameter of the field component must turn from +1 to $\simeq -1$.
This must occur {\it about a suitable redshift $z_\pm$} so to allow a
fair amount of today's DE. In any approach based on a self interaction
potential $V(\Phi)$, the fair $z_\pm$ value is obtained by tuning some
parameter(s) inside $V(\Phi)$ expression itself. SCDEW models behave
similarly. It is just convenient to make use of the first order field
equation
\begin{equation}
\dot \Phi_1 + \tilde w {\dot a \over a} \Phi_1 = {1 + w \over 2}
C\rho_c a^2~,
\end{equation}
instead of eq.~(\ref{eq6}); here $\Phi_1 \equiv \dot \Phi$ and $2
\tilde w = 1+3w-d\log (1+w)/d\log a$. This equation, shown by BSLV12,
requires $w(a)$ to be assigned, instead of a potential $V(\Phi)$.  In
this way, we select $z_\pm$ directly, without arguing about untestable
$V(\Phi)$ expressions. More explicitly, let us assume that
\begin{equation}
w = {1-A \over 1+A} ~~~ {\rm with} ~~~ A = \left( a \over a_\pm
\right)^\epsilon
\label{transition}
\end{equation}
with $a_\pm= (1+z_\pm)^{-1}$. In BM14 we also discussed the (minimal)
dependence of results on the parameter $\epsilon$ ($=2.9$ here), whose
arbitrariness replaces the choice of $V(\Phi)$ expression.

An option, not mentioned by BSLV12 and BM14, is that $\beta$ gradually
fades, after the rise of $\rho_{\rm w}$ broke conformal invariance. No
wanted feature of this class of cosmologies really depends on a late
coupling and we shall see that, if $\beta$ fading occurs late enough,
just minor quantitative changes arise.

\begin{figure}
\begin{center}
\includegraphics[width = 0.4\textwidth]{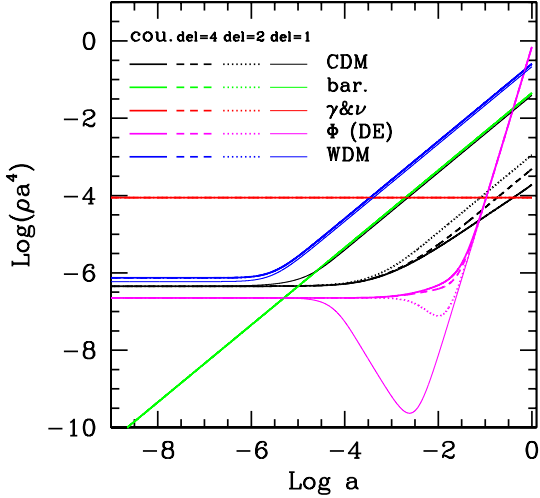}
\end{center}
\caption{Scale dependence of the densities of cosmic components after
  the break of conformal invariance. All models in this plot have
  $\beta=10$, $m_{\rm w}=90\, $eV. Delay ($\rm del$) values from 1 to $\infty$
  (ever lasting coupling) are considered. The thick curves labelled
  ``cou'' concern an ever lasting coupling. The behavior of densities
  does not suffer major changes, in respect to this option, unless
  $d < 2$. }
\label{den}
\end{figure}

Let us then consider the option that $\beta \propto \exp(-a/a_{dg})$
with $a_{dg} = D \times a_{\rm w}$. If we expect $\beta$ fading to be
a sort of {\it consequence} of the loss of conformal invariance,
if should be $D \gg 1$. Accordingly, different options will be
labelled by the value of the delay (del.)
\begin{equation}
\label{delay}
d = \log_{10} D
\end{equation}
which is an extra parameter we introduce here. In Figure \ref{den} we
show the scale dependence of cosmic components in 4 different cases,
passing from the option of ever lasting coupling, down to the case of
delay $d = 1$. In this Figure and anywhere in paper I, we select the
following parameter values:
$$
\begin{matrix}
\Omega_{0\Phi} & \Omega_{0b} & h_0  &  T_{CMB} & N_\nu & n_s & \cr
 0.7  & 0.045 & 0.685 & 2.726 & 3.04  & 0.968 & \cr
\end{matrix}
$$ Here $\Omega_{0\Phi}$, $ \Omega_{0b}$, $ h_0$, $ T_{CMB}$, $ N_\nu$, $
n_s $ are the present DE and baryon density parameters, the present
Hubble parameter in units of 100 (km/s)/Mpc, the CMB temperature, the
number of (almost) massless $\nu$'s, the primeval scalar fluctuation
index, respectively. In the list, $\Omega_{0{\rm w}}$ and
$\Omega_{0c}$ are not included. Neglecting $\gamma$'s and $\nu$'s,
their sum is $1-\Omega_{0\Phi}-\Omega_{0b}$, while $\Omega_{0,c}$ is
fixed by the selection of the coupling constant $\beta$ and
$\Omega_{0{\rm w}}$ covers the residual gap. Notice that, at $z=0$,
$\Omega_{0,c} \ll 1/2\beta^2$ being, typically, $\sim 10^{-2}
\Omega_{0b}$ (see Figure~\ref{den}).
 
The option $d = 4$, that will be selected in Paper II, is
characterized by $a_{dg} \simeq a_\pm$; as shown in Figure \ref{den},
for $d = 4$ density evolusion is just marginally affected by the
fading of coupling.

An earlier decoupling, instead, bears a number of consequences. Among
them, a rise of today's CDM density parameter $\Omega_{c,0}$, because
the energy leaking from CDM to DE has a stop. If keeping $d \gtrsim 2$,
this rise is limited and, in this range of $d$ values, the present CDM
density parameter $\Omega_{c,0} \simeq 10^{-2} \times \Omega_{b,0} $:
the present CDM contribution to the cosmic budget is $\sim 1/100$ of
baryons.

A related effect is the formation of a dip in $\rho_\Phi a^4$
evolution. As a matter of fact, until $\Phi$ is purely kinetic, its
density keeps $\rho_\Phi = m_p^2/(2 b\, a^2 \tau^2)$, thanks to the
energy inflow from CDM. In the radiative era, when $\tau \propto a$,
$\rho_\Phi a^4$ is then constant. The transition to matter dominance
should strengthen $\rho_\Phi$ decline; however, when this occurs,
coupled CDM density starts to increase, so providing a stronger $\Phi$
feeding.
If the feeding stops before the kinetic--potential transition,
occurring at $z_\pm$, a progressive dilution of $\rho_\Phi$ is
unavoidable. However, this produces a significant dip only if $d <
2$.  Owing to these reasons, in the rest of this Paper I we shall
never consider options $d < 2$. We shall also verify that models with
any $d > 2$ exhibit just minor differences.

All through this discussion, we deliberately refrained from
  introducing any detailed physics causing a $\beta$ fading. Possible
  mechanisms are discussed in Appendix A and further options surely
  exist, but the introduction of a further parameter can hardly be
  avoided. The treatment given in this work keeps on the
  phenomenological side, just showing that $\beta$ fading can be self
  consistently introduced, so allowing a plain treatment of nonlinear
  stages.

Early nonlinear evolution will can be expected to induce a
hierarchical rippled CDM distribution, made of ``virialized''
structures, on growing scales.

When and if $\beta$ fades, CDM 
turns into a component of fast heavy particles. We expect them to
cause no substantial effect on the linear evolution of other
components, also because of the smallness of the CDM density
parameter; on this ground we decided to not consider these effects in 
our linear treatment.

Before concluding this discussion on background evolution let us
outline that the relation between expansion factor $a$ (or redshift)
and ordinary time $t$ are almost identical in SCDEW and $\Lambda$CDM
models; discrepancies never exceed 0.01\% for any values of coupling
$\beta$ and delay $d$.

\section{Linear fluctuation evolution}\label{sec:Lin-evol}

In BM14 fluctuation evolution is studied in detail.  Out of horizon
{\it initial conditions} are determined and the system of differential
equations, holding after the entry into the horizon, is numerically
solved. Here, let us only remind some peculiar aspects, that this
problem does not share with similar cosmological models. In
particular, field fluctuations are conveniently described by a
variable $\varphi$, related to the quintessential field $\phi$ as
follows:
\begin{equation}
\phi = \Phi + {b \over m_p} \varphi~,
\label{varphi}
\end{equation}
$\Phi$ being the background field. At the first perturbative order
$\varphi$ fulfills the equation
\begin{equation}
\ddot \varphi + 2{\dot a \over a} \dot \varphi + \dot \Phi \dot h + k^2
\varphi + a^2 V''(\Phi) \varphi = 2 \beta^2 \Omega_c \left( \dot a
\over a \right)^2 \delta_c~.
\label{phieq}
\end{equation}
Derivatives are taken in respect to the conformal time, as the metric
reads
\begin{equation}
ds^2 = a^2(\tau) [d\tau^2 - (\delta_{ij}+h_{ij}) dx_i dx_j)]~.
\end{equation}
Moreover, $\delta_c$ are the fluctuations in the coupled CDM component
and $h$, the trace of the 3--tensor $h_{ij}$, describes gravity. 
Finally, $k$ yields the mass scale $ M = (4 \pi / 3) \rho
\left( 2\pi / k \right)^3 $ of the fluctuation.  

Most terms in eq.~(\ref{phieq}) have therefore a transparent meaning,
apart of the term $V''(\Phi)$, where we apparently refer to a $\phi$
self--interaction potential. Aiming to skip any reference to such
quantity, BM14 show that one can use the relation
$$
2V'' = {A \over 1+A} \bigg\{ {\dot a \over a} {\epsilon \over 1+A}
\left[ \epsilon_6 {\dot a \over a^3} + 2C {\rho_c \over \dot \Phi}
\right] + $$ 
\begin{equation}
+ \left[ {\dot a \over a^3} {\ddot \Phi \over \dot \Phi}
+ {d \over d\tau} \left( \dot a \over a^3 \right) \right]
\epsilon_6 + 2C {\dot \rho_c \over \dot \Phi} \bigg\}
\end{equation}
with $\epsilon_6 = \epsilon-6$. Here, $A$ and $\epsilon$ are the
quantities defining the $w(a)$ behavior in eq.~(\ref{transition}).
All variables in this expression are known at each step of a linear
evolutionary algorithm. 

It is then significant that, within this formulation, also in the
fluctuation equations only $\dot \Phi$ and $\ddot \Phi$ appear, while
$\Phi$ values never matter.
\begin{figure}
\begin{center}
\includegraphics[width = 0.4\textwidth]{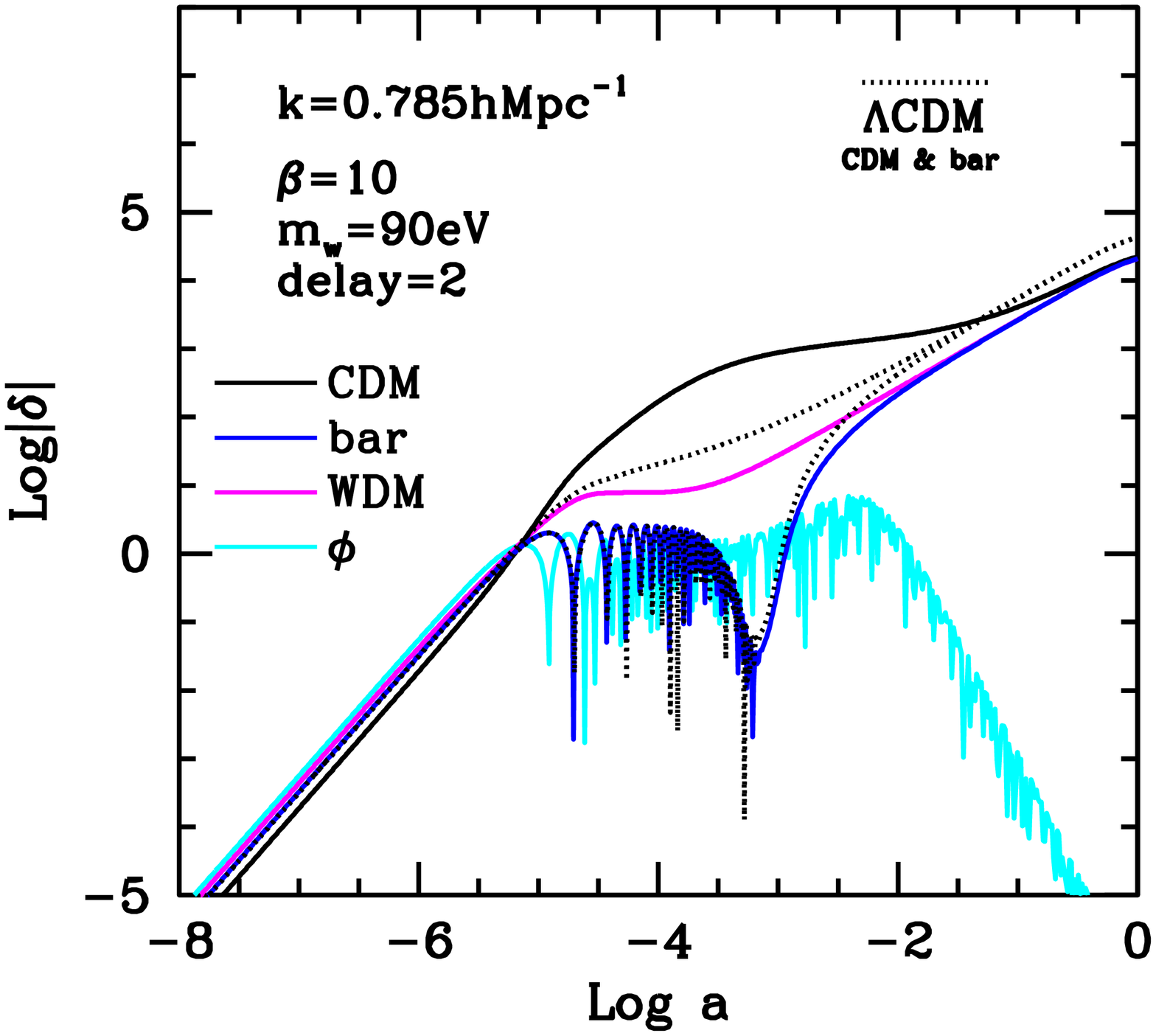}
\vskip -.4truecm
\includegraphics[width = 0.4\textwidth]{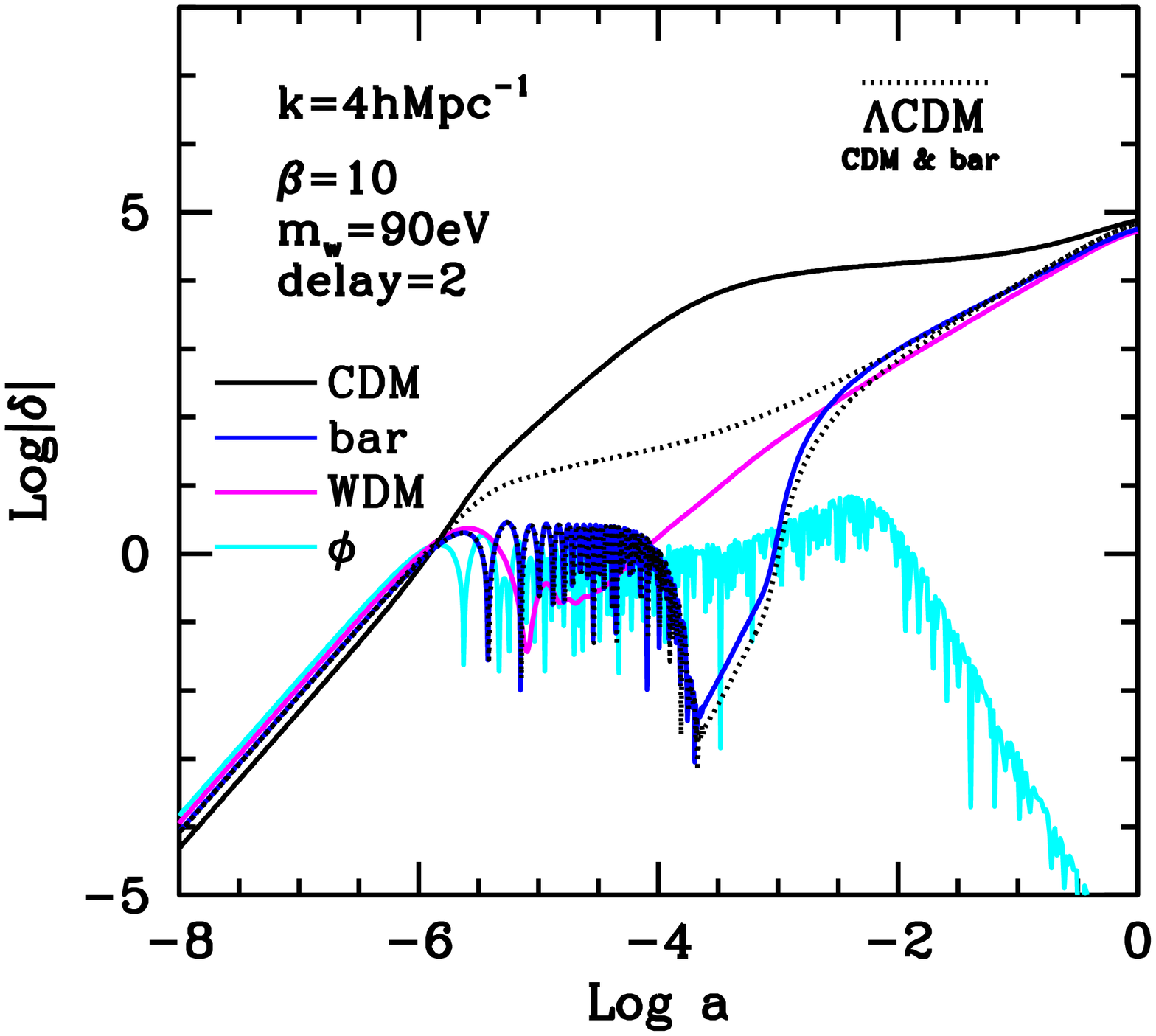}
\vskip -.4truecm
\includegraphics[width = 0.4\textwidth]{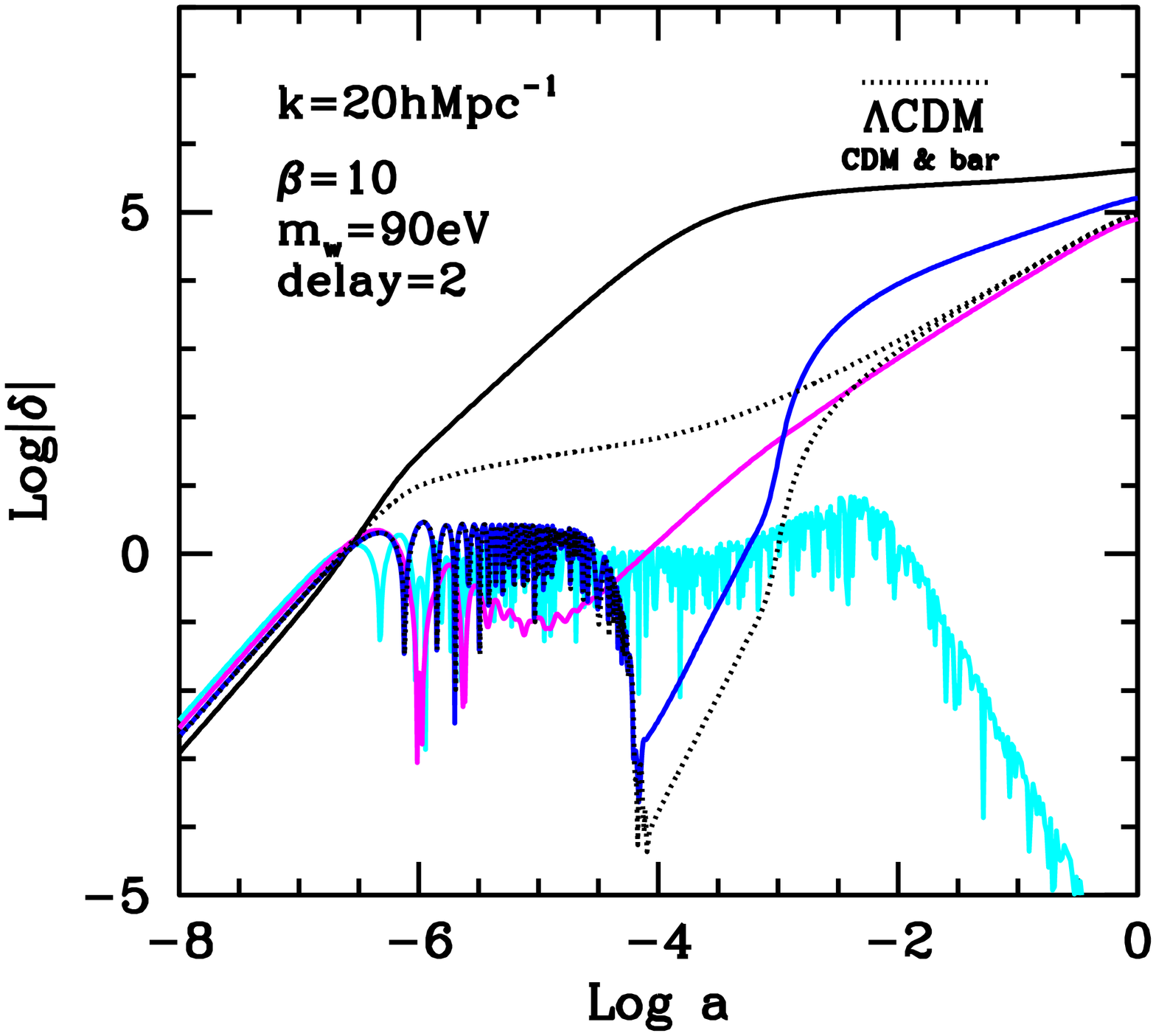}
\end{center}
\caption{Evolution of density fluctuations in the model indicated in
  the frame for 3 $k$ values; from top to bottom they correspond to
  masses $\sim 1.8 \times 10^{14}$--$~ 1.35 \times 10^{12}$--$~ 8.7
  \times 10^{10} h^{-1}M_\odot$, respectively. The first $k$ value
  then corresponds to a comoving length scale of $8\, h^{-1}$Mpc, at
  the boundaries between linear and non--linear behaviors; the second
  and third $k$ values lay at the top and bottom limits of the
  galactic mass scales. Dotted curves show the evolution of CDM and
  baryons, for the same scales, in a $\Lambda$CDM model.  }
\vskip -.3truecm
\label{evolve}
\end{figure}
\begin{figure}
\begin{center}
\includegraphics[width = 0.4\textwidth]{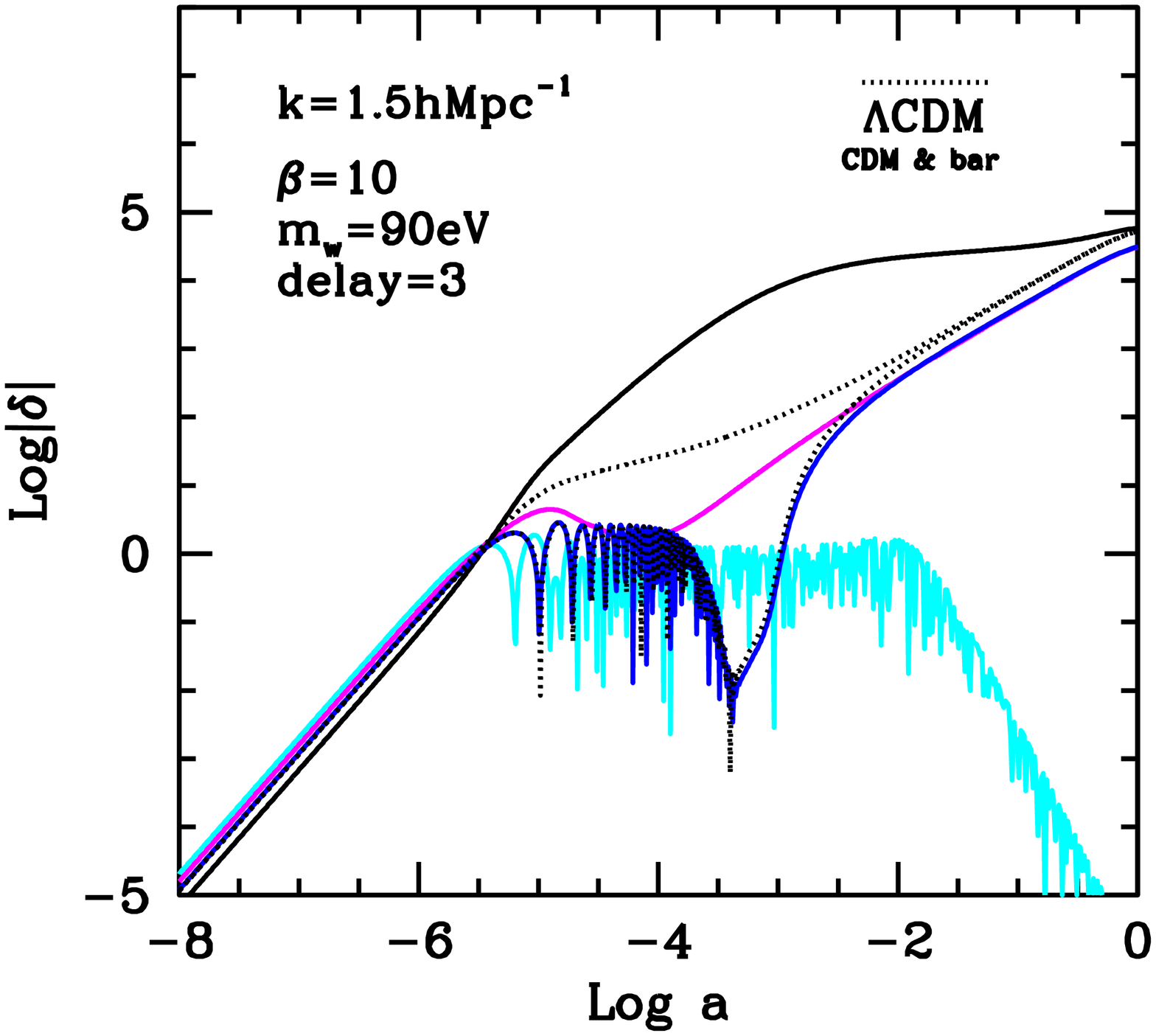}
\vskip -.4truecm
\includegraphics[width = 0.4\textwidth]{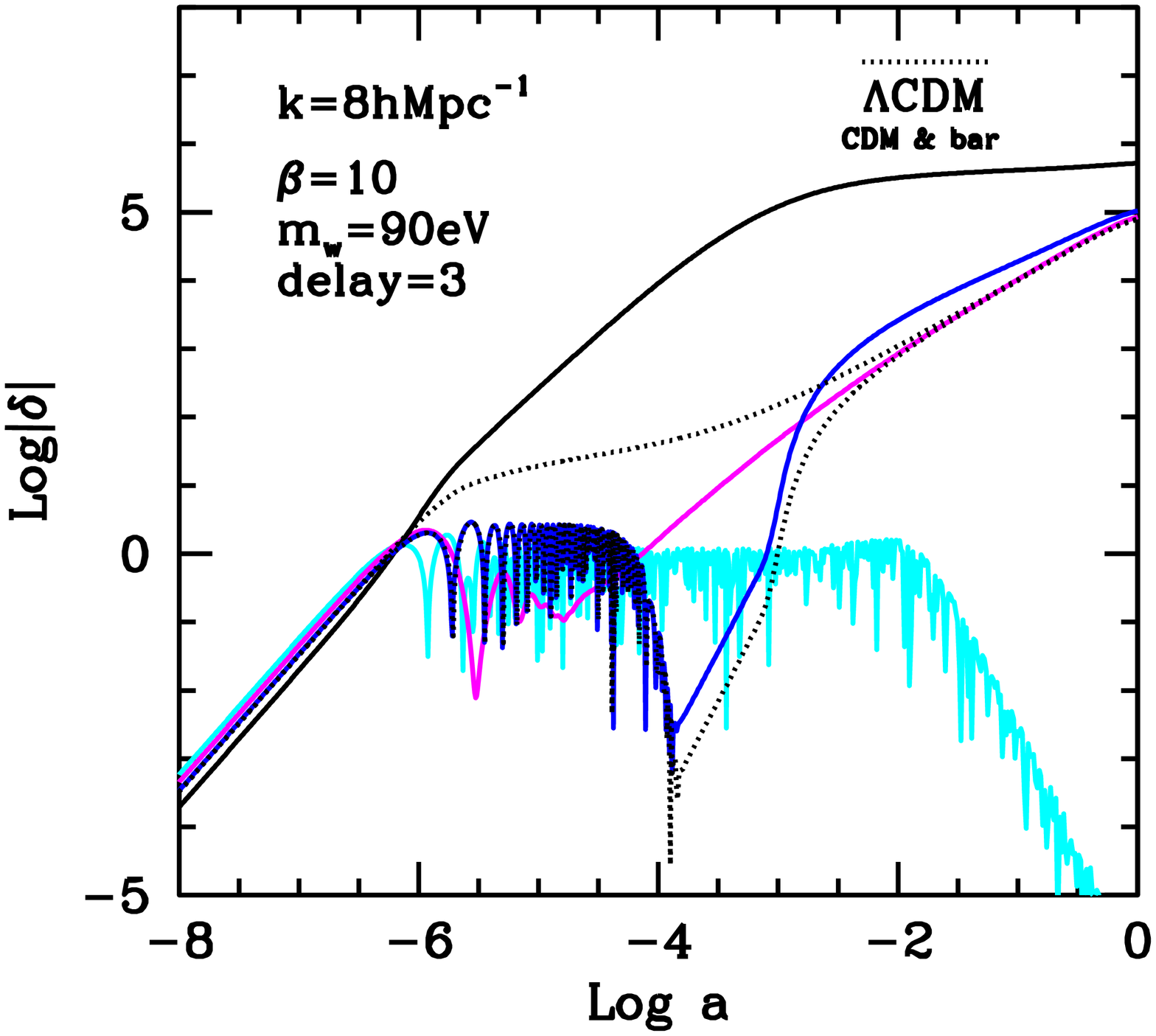}
\end{center}
\caption{As previous Figure, for ${d}=3$, and 2 intermediate scales. }
\label{evolve1}
\end{figure}
BM14 made use of these relations to modify the public algorithm {\sc
  cmbfast} and here we shall report results obtained with this code.

\subsection{Early growth of coupled CDM fluctuations}

Figure \ref{evolve} and \ref{evolve1} then show a few examples of
linear evolution, for fluctuations in the different cosmic components,
from before their entry in the horizon, down to $z=0$. The background
cosmic parameters are the same of Figure \ref{den}. The mass of the
warm component is anywhere 90$\, $eV, its range being fixed by the
arguments discussed below, in the next subsection. The $k$ values
selected for Figure \ref{evolve}, where the delay parameter is $d=2$,
correspond to the mass scales of $1.8 \times 10^{14} \hMsun$ (or 8
\Mpch, top panel), $1.4 \times 10^{12} \hMsun$ (central panel), and
$8.7 \times 10^{10} \hMsun$ (lower panel).  Figure \ref{evolve1} is
for 2 intermediate scales and $d=3$. For the sake of comparison, the
evolution of baryon and CDM fluctuations in a $\Lambda$CDM model with
the same cosmological parameters is also overplotted.

A first result is shown in the top Figure \ref{evolve}: the low--$z$
growth of fluctuations, namely on (scales close to) the linear regime,
is the same in $\Lambda$CDM and SCDEW cosmologies. Numerical outputs
confirm that discrepancies between $\Lambda$CDM and SCDEW growth
factors, for $k \lesssim 0.5$, are below the expected precision of the
algorithms used. It is so in spite of the (slightly) different final
amplitudes. In the non--linear regime, discrepancies are appreciable,
although still small. Aside of that, it is worth outlining that SCDEW
models are also indistinguishable from $\Lambda$CDM, as far as CMB
spectra are concerned. Discrepancies between SCDEW and $\Lambda$CDM
CMB spectra, for $\beta \gtrsim 5$ are again below the expected
precision of the algorithms used.

Another evident feature is the fast growth of {\it coupled} CDM
fluctuations, not discontinued at the time when the horizon attains
the fluctuation size. In $\Lambda$CDM, {\it uncoupled} CDM fluctuation
grow much more slowly, namely until radiation exceeds CDM density, as
radiation fluctuations (still coupled to baryons) have stopped their
growth, being in the sonic regime. In SCDEW models, this extra growth
is expected, in spite of the small density $\rho_c \simeq
\rho_t/2\beta^2$ ($\rho_t:$ total density) of the coupled CDM
component: in the non--relativistic regime, coupling causes an
increase of the {\it effective} CDM--CDM gravity by a factor
$1+(4/3)\beta^2$ \citep{amendola1999}; {\bf for $\beta^2 \gg 3/4$,}
  this factor erases the division by $2\beta^2$ so that, as far as
  self--gravity is concerned, it is as thought CDM had a density close
  to the total density $ \rho_t$.
 In addition, CDM particle velocities are enhanced by a continuous
  push due to their progressive mass decrease (the so--called antidrag
  term). All these effects are taken into account by the linear
  program and cause an increase of CDM density fluctuations, after
  their entry in the horizon and in the radiation dominated epoch,
  approximately~$\propto a^{3/2}$.


\subsection{WDM fluctuation regeneration}

Large amplitude CDM waves cause a restart of fluctuations in WDM, as
soon as it becomes non--relativistic. As is also visible in Figures
\ref{evolve} and \ref{evolve1}, WDM derelativizes at $z_{{\rm w},der}
\sim 10^5$. Derelativization is linked to the value taken by the
$T_{\rm w}/T_\gamma$ ratio, between WDM and photon temperatures. If
WDM particles decoupled from the {\it thermal soup} when its density
was $(\pi^2/30) g_{{\rm w},dec} T^4$, today $(T_{\rm w}/T_\gamma)^3
\simeq (2/g_{{\rm w},dec}) (1 + 3N_\nu/22)$ ($N_\nu:$ number of
(almost) massless $\nu$'s in the cosmic background; $g_{{\rm w},dec}$:
number of relativistic degrees of freedom at the time of WDM decoupling), while
\begin{equation}
\Omega_{0{\rm w}} h^2 \simeq {45 \zeta(3) \over 4\pi^4} \left(T_{\rm w}
\over T_\gamma \right)^3 {m_{\rm w} \over 10^4 T_\gamma}~.
\end{equation}
For $N_\nu = 3$ we have then
\begin{equation}
\Omega_{0{\rm w}} h^2 \simeq 0.115 {m_{\rm w}/{\rm eV} \over g_{{\rm w},dec}}
\end{equation}
so that $\, T_{\rm w}/T_\gamma \sim 0.32$ if $\Omega_{{\rm w},0} h^2
\simeq 0.12 $ and $m_{\rm w }\simeq 90~$eV. Accordingly, $T_{\rm w} \simeq
m_{\rm w}$ at a redshift $\sim 10^5$--$10^6\, $, in agreement with
what is shown in the plots, these effects being all included in
modified {\sc cmbfast}. In particular, our algorithm takes into
account the contributions of coupled CDM and $\Phi$ field to the {\it
  thermal soup}.

At the time of WDM derelativization, CDM fluctuations $\delta_c$
exceed then any other fluctuations, by a scale dependent factor
$F_{b} \sim 10^2$--$10^3$ for galactic scales. There is no boost to
the CDM--WDM gravity due to coupling, but the CDM density excess
$\delta \rho_c = \rho_c \delta_c \simeq (\rho_t/2\beta^2) \delta_c$ is
boosted by the factor $F_{b}$, compensating the division by
$2\beta^2 \sim 10^2$ due to the small CDM density. As the factor
$F_{b}$ depends on the time elapsed since the entry in the horizon,
the restart of WDM fluctuations should be more and more effective
towards greater $k$ values. In turn, WDM particle escape velocity from
smaller size fluctuations is smaller, so that the effect is only
partially visible in the final spectra.

On the contrary, this very effect is unchallenged in baryons, whose
mean velocities are negligible. Also the (later) restart of their
fluctuations is mostly due to CDM. In the absence of residual particle
velocities, baryon fluctuations growth  even exceeds WDM
fluctuations, both at high $z$ and at $z=0$, namely at high $k$.

Since $\sim z_{{\rm w},der}$, WDM density starts to dilute just
$\propto a^{-3}$, and primeval conformal invariance is broken. Also
baryon density growth violates such invariance, of course, but baryon
density overcomes radiative components at a redshift $ < z_{{\rm w},der}$.
In principle, it can make sense that a break of
conformal invariance and $\beta$ fading are related events: when WDM
has turned non relativistic, the $\beta$ coupling appears like a
residual pleonasm.  Leaving apart detailed options to model a relation
between these effects, we just quantify it with the parameter defined
in eq.~(\ref{delay}). For instance, for $d = 2$ $(3,\, 4)$, $\beta $
fades when $z \sim 5000$ $(500,\, 50)$. Values of $d \gg 6$ return an
ever lasting $\beta$--coupling.

\subsection{WDM particle mass selection}

WDM fluctuation regeneration in SCDEW models contrasts with undisputed
free streaming effects in standard $\Lambda$WDM models, causing a
large--$k$ cutoff to the spectral function
\begin{equation}
\Delta^2(k) = {1 \over 2\pi^2} k^3 P(k)~,
\label{d2}
\end{equation}
($P(k) = \langle | \delta(k) |^2 \rangle:$ transfered spectrum),
because of the erasing of any fluctuation entering the horizon before
WDM has derelativized. With $m_{\rm w} \simeq 90\, $eV, the decline of
$\Delta^2$ starts at $k \simeq 0.5\, h$Mpc$^{-1}$, $\Delta^2$ being
already damped by a factor $\sim 10^3$ at $k \simeq 1\, h$Mpc$^{-1}$.
Accordingly, viable $\Lambda$WDM models currently refer to WDM masses
$m_{\rm w} \simeq 2$--3~keV, allowing fluctuations to survive up to $k
\simeq 20$--$30\, h$Mpc$^{-1}$.

Unfortunately, the need of such a large $m_{\rm w}$ partially invalidates
the choice of warm instead of cold DM. The residual motions of
low--mass particles, e.g., are then insufficient to reduce the number
of expected MW satellites or to prevent them settling on a NFW
(Navarro, Frank \& White 1997) profile. Recent simulations confirm a
relation between $m_{\rm w}$ and the size of a core. According to
\citet{maccio2012}, in the dwarf galaxy mass range, the size of the
core
\begin{equation}
R_{core} \sim   \left( {1.0} \over {\hkpc} \right)   \left( 100\, 
{\rm eV} \over m_{\rm w}
\right)^{1.8}~.
\label{rcore}
\end{equation}
If $m_{\rm w} \sim 2\, $keV, e.g., it is $r_{core} \sim 5\,
h^{-1}$pc$\, $.  On the contrary, a fair core size, fitting
observations, apparently requires $m_{\rm w} \sim 80$--110$\, $eV; a
mass scale yielding fair cores but no galaxies. It is then
significant that the linear theory of strongly coupled--DE models
exhibits a full restart of fluctuations.

Before concluding that a solution to low--scale $\Lambda$CDM problems
is found, however, we shall first consider in detail the full shape of
linear spectra and use them to perform {\it ad--hoc} numerical
simulations. The former aim is fulfilled here below, the latter one is
the target of Paper II.

Meanwhile, let us outline that there are a few other elements in favor
of choosing $m_{\rm w} \simeq 90\, $eV, indipendently of the expected
halo and satellite predictions. The very Figure~\ref{den} shows that
the contributions of WDM, CDM and $\Phi$ to the primeval {\it thermal
  soup} are all within half order of magnitude. Even though this is
due to the choice of $\beta$ and $m_{\rm w}$, it would be reckless to
conclude that it fixes their ranges, but it would also be hard to
believe all that to be fully casual. 
\begin{figure}
\begin{center}
\vskip -1.3truecm
\includegraphics[width = 0.47\textwidth]{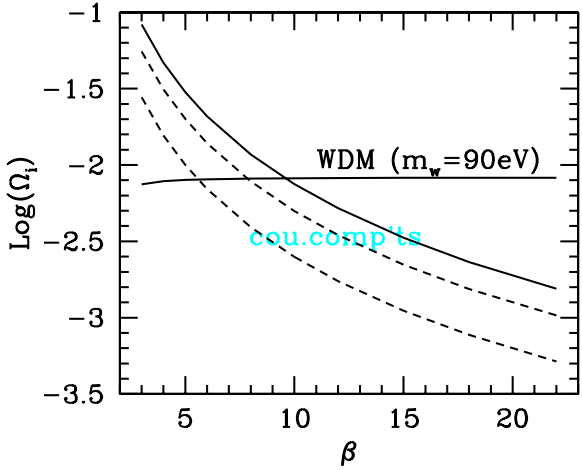}
\end{center}
\caption{Primeval density parameter of WDM and coupled CDM--$\Phi$
  components in the radiation dominated era, for $m_{\rm w} = 90\,
  $eV. The upper (lower) dashed line yields CDM ($\Phi$) density. The
  total density of the coupled components is the ``parallel'' solid
  curve. The slightly rising solid line is WDM density parameter.}
\label{denso}
\end{figure}

Figure \ref{denso} then shows a strictly related point: The primeval
density parameters $\Omega_i$ plotted there are for CDM, $\Phi$ and
WDM. The WDM density parameter exhibits just quite a mild $\beta$
dependence. In fact, only for very low $\beta$ values, its level is
significantly eaten by a non--negligible $3/4\beta^2$ contribute.  On
the contrary, the WDM level would siginificantly depend on $m_{\rm
  w}$.  Should this mass be greater by $\Delta m$, WDM derelativize
earlier by $\Delta a \simeq \Delta m /T_0$ and the WDM line should be
lower. In order to recover the nearly--coincidence between primeval
$\Omega_i$, fairly greater $\beta$ values should then be selected, as
the dependence on $\beta $ is quadratic. The opposite would occur for
lower $m_{\rm w}$ values, with the risk to approach the limits of the
physical range for~$\beta$'s. 

This plot should be however taken together with a third argument,
which can be made only after giving details on spectra (see Figure
\ref{sigbet}, below): equal normalizations for CMB angular spectra and
$\sigma_8$, in $\Lambda$CDM and SCDEW models, should be required; this
fixes a narrow $\beta$ range, once $m_{\rm w} = 90\, $eV is selected.
In turn, Figure \ref{denso} returns $m_{\rm w} = 90\, $eV for that
$\beta$ range. Altogether, the two constraints are close to a system
of 2 algebric equations with 2 unknown, although yielding softer
constraints. $\beta \simeq 10 $ and $m_{\rm w} = 90\, $eV are at the
middle of the narrow allowed bands. If N--body simulations will
confirm that this selection favours observational profiles, etc., we
may say that this follows from the optimal parameter choice within the
model.

\subsection{CDM non--linearities}
\begin{figure*}
\begin{center}
\includegraphics[width = 0.45\textwidth]{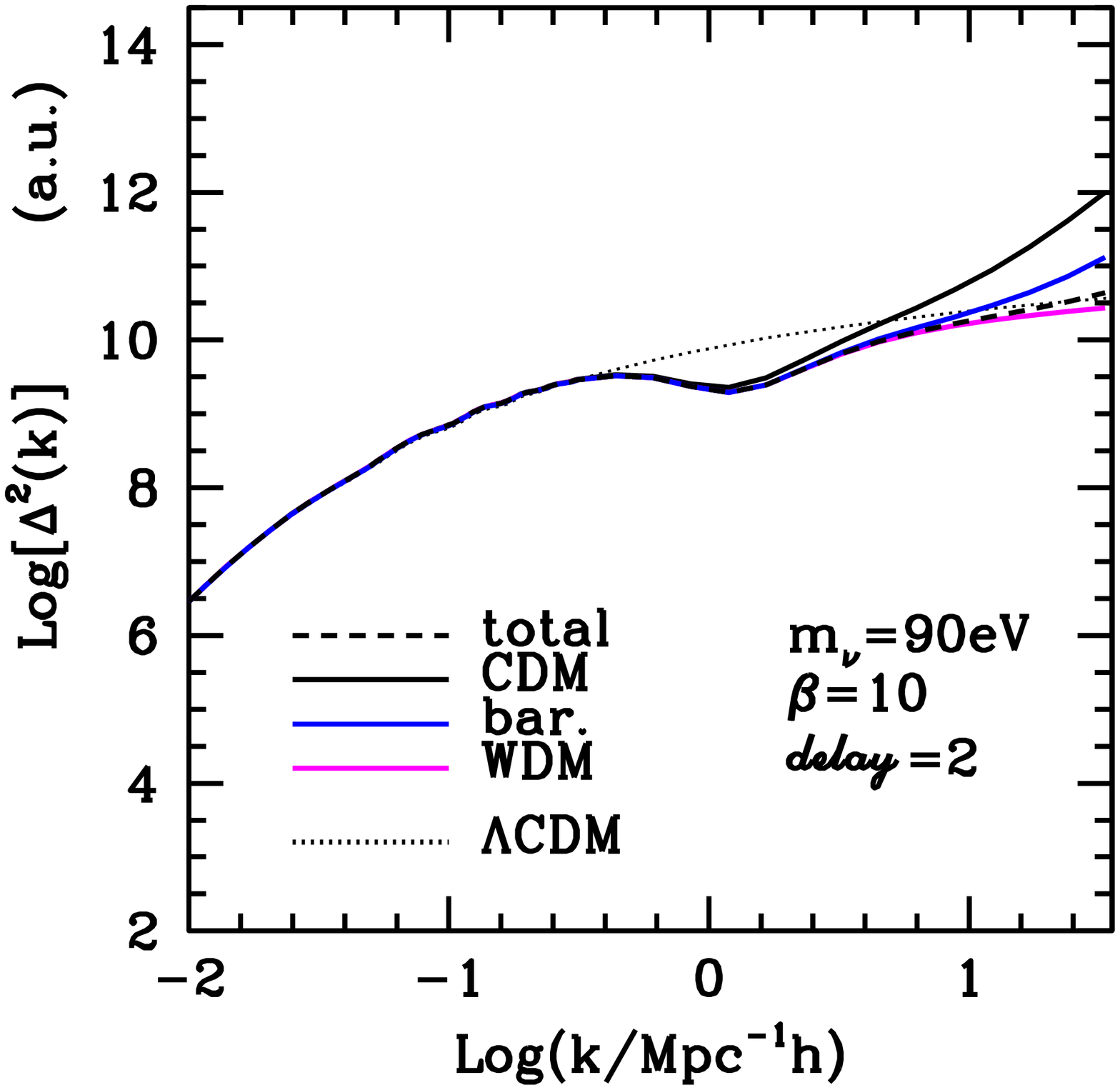}\hspace{12mm}\includegraphics[width = 0.45\textwidth]{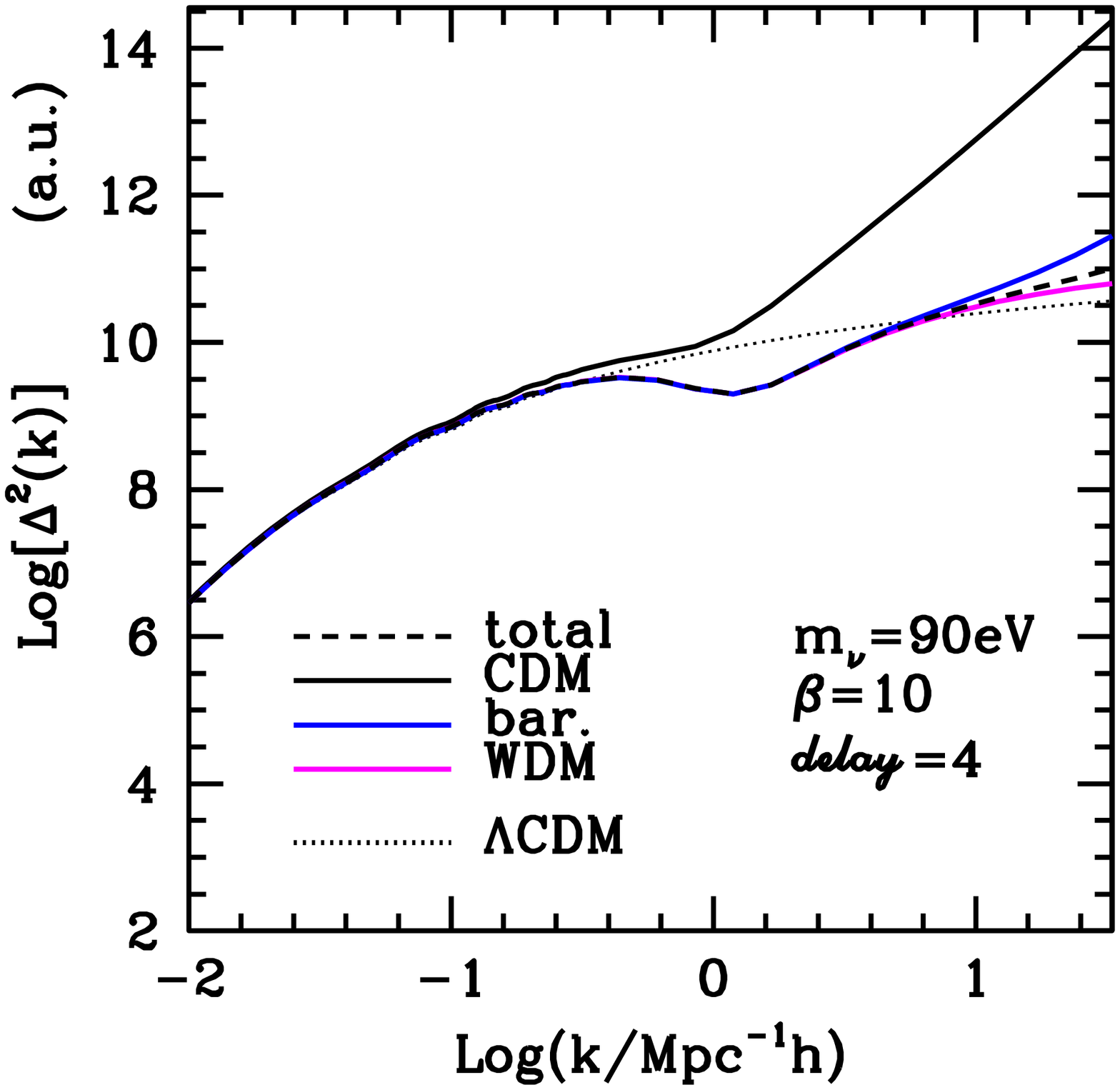}

\end{center}
\caption{Spectral functions obtained from the linear theory at $z=0$
  for models with $d=2$ and 4. We compare them with $\Lambda$CDM
  spectra, finding that the {\it revival} of high--$k$ fluctuations is
  highly efficient.  }
\label{tfa}
\end{figure*}

Recovering (WDM or) baryon fluctuations thanks to coupled CDM action,
on those scales where they had been erased, requires CDM fluctuations
100--1000 times wider than uncoupled CDM fluctuations in $\Lambda$CDM
models. In fact, coupled CDM has a much smaller density parameter than
{\it usual} CDM, when its action is needed.  Luckily enough, such wide
amplitude is an intrinsical feature of SCDEW models. In turn, this
feature can cause the birth of early CDM nonlinearities.
These non--linearities are anyway suppressed by the fact that CDM 
contributes to the cosmic budget as little as $\sim 1$ per cent of 
the very baryonic component.

In SCDEW models early non linear CDM structures however form on small
scales. In the radiation dominated period, after entering the horizon,
coupled CDM fluctuations grow $\propto \sim a^{3/2}$.  If $\delta_c
\sim 10^{-5}$ at the horizon redshift $z_h$, it shall be $\delta_c
\sim 1$ at $z_{n.l.} \sim z_h \times 10^{-10/3}$. Accordingly,
fluctuations on scales entering the horizon at a redshift $\sim 10^3$
times greater than equality, involving a mass $\lesssim 10^7 h^{-1}
M_\odot$ are quite likely to produce CDM non linear structures. Higher
mass structures can also form up to a scale depending on the delay
$d$.

The most immediate problem related to CDM non--linearities concerns the
reliability of linear algorithms predicting fluctuation evolution.  In
fact, the first effect due to non linearity onset, is an acceleration
of the growth rate. An estimate of the effect can be desumed from the
theory of spherical growth in pure CDM models, telling us that a
density contrast $\simeq 5$--6 is attained when the linear $\delta_c
\simeq 1.0$--1.1~. Non--linearity triggers other effects, as mode
mixing, a loss of Gaussianety and high velocity fields. Forming
structures, in particular, shall have a shrinking radius and a global
motion, so that we can expect a substantial loss of coherence between
the distribution of CDM and other cosmic components, with top spectral
contribution eventually affecting wavelengths smaller than the
original fluctuations $\delta_c$. Accordingly, no excessive effects
are expected on scales never overcoming a mild non--linearity.

At an advanced stage, evolved CDM non--linearities shall then be
non--linear structures embedded in an almost unperturbed continuum,
taking also into account that, altogether, the CDM density parameter
has reduced to $\cal O$$(10^{-3}$--$10^{-4})$, so that the perturbing
bodies shall be relatively rare.

Let us also outline that, also after $\beta$ fading, the linear
algorithm will continue to treat CDM and baryons separately.  As a
consequence, the CDM fluctuation $\delta_c$ can still overcome unity
while the baryon fluctuation $\delta_b$ is still $\ll 1$. However,
in these conditions, CDM and baryon obey the same equations of
motion and the physical variable is 
\begin{equation}
\delta_{cb} = {\Omega_c \delta_c + \Omega_b \delta_b \over \Omega_c +
  \Omega_b}~.
\end{equation}
Unless this is non--linear, physical non--linearities do not exist.
Also if, slightly before $\beta$ fading, $\delta_c$ approached unity,
but without badly loosing coherence, the physical post--fading
variable will be $\delta_{cb}$; however, even if coherence is
partially lost, that part of $\delta_c$ still coherent with $\delta_b$
will enter a unified growth regime. Accordingly, the effect of
marginal non linearities can be thought to be just a slight boost of
the CDM+baryon fluctuation amplitudes. A quantitative estimate of such
boost is not immediately feasible, although one must however recall
that $\Omega_c \ll \Omega_b$, so that the very uncertainty shall not
be too large.

When considering the spectral function in the next Section, which are
a product of linear algorithms, one must therefore discriminate among
different possible {\it meanings} of apparently high $\Delta^2(k)$ for
the CDM component at high $k$.

There is however a clear conclusion for the above discussion: SCDEW
models predict the formation of early non--linear CDM structures up to
a mass $\sim 10^7$--$10^8 h^{-1} M_\odot$. With a present CDM density
$\sim 10^{8} h^{-1}M_\odot/(h^{-1}{\rm Mpc})^3$, we expect their
average comoving distance to be $\sim 1\, h^{-1}$Mpc, a value
comparable with the average intergalactic distance.
Smaller residual CDM structures could also exist, if not embedded in
greater structures in a hierarchical formation process. 
No further speculation is however possible, to better define the
specific features these ``objects'' could exhibit, apart of the rather
obvious but probably simplistic statement that they might occupy the
nucleus of existing galaxies.

\begin{figure}
\begin{center}
\vskip -0.8 truecm
\includegraphics[width = 0.45\textwidth]{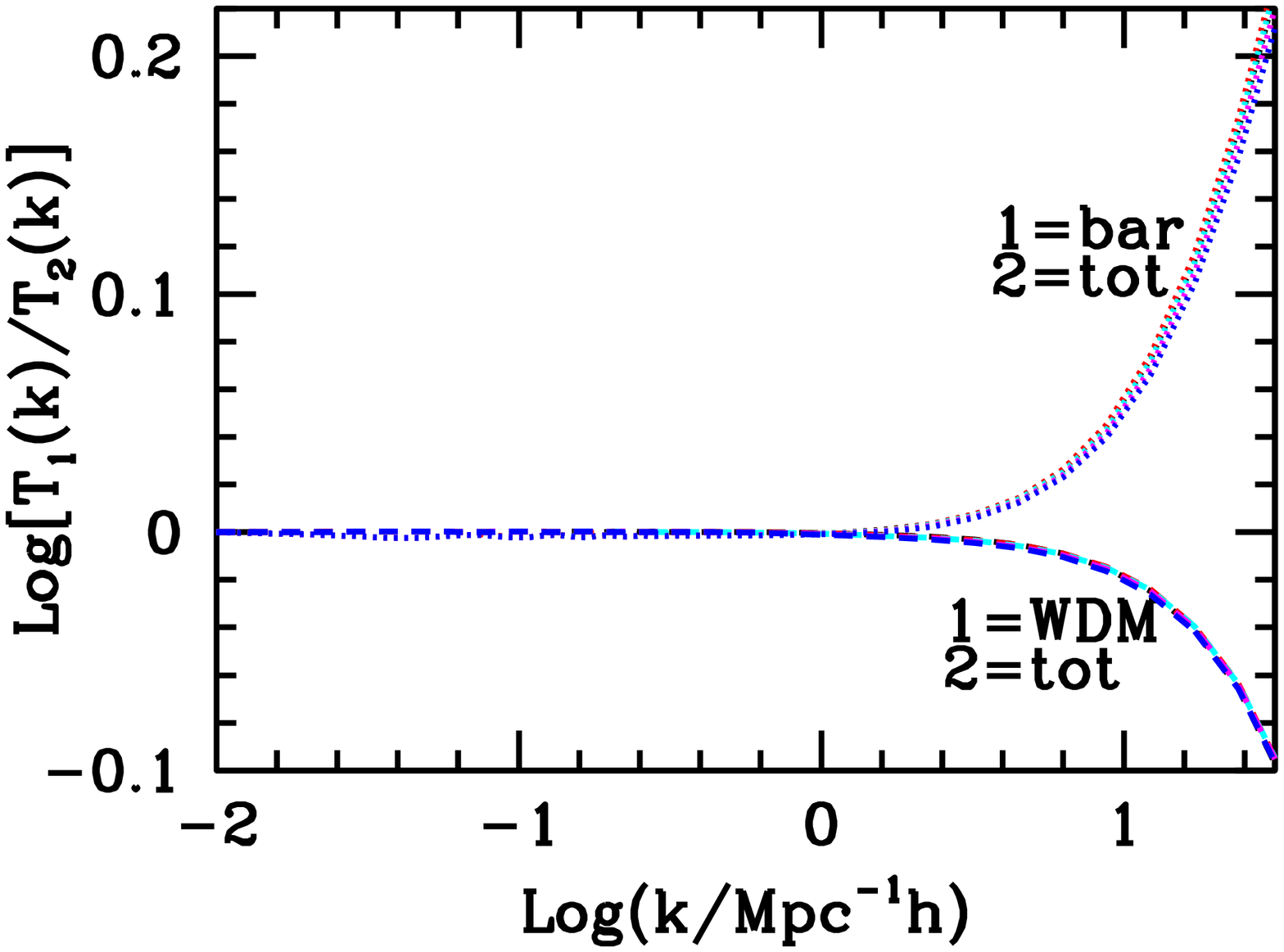}
\end{center}
\vskip -0.6 truecm
\caption{High--$k$ difference between baryon and WDM transfer
  functions T(k), shown to be independent from $d$ (we overlap curves
  for $d= 1.8,~2.4,~3,~3.5,~4$). The ratios ${\rm T}_1/{\rm T}_2$ are
  between the transfer functions for baryon or WDM and the total
  transfer function, as indicated in the Figure. The model is $m_{\rm
    w}=90\, $eV and $\beta=10$. As expected, the split starts at
  greater (smaller) $k$ if a greater (smaller) $m_{\rm w}$ is taken. }
\label{wb}
\end{figure}
\begin{figure}
\begin{center}
\vskip -1.7truecm
\includegraphics[width = 0.54\textwidth]{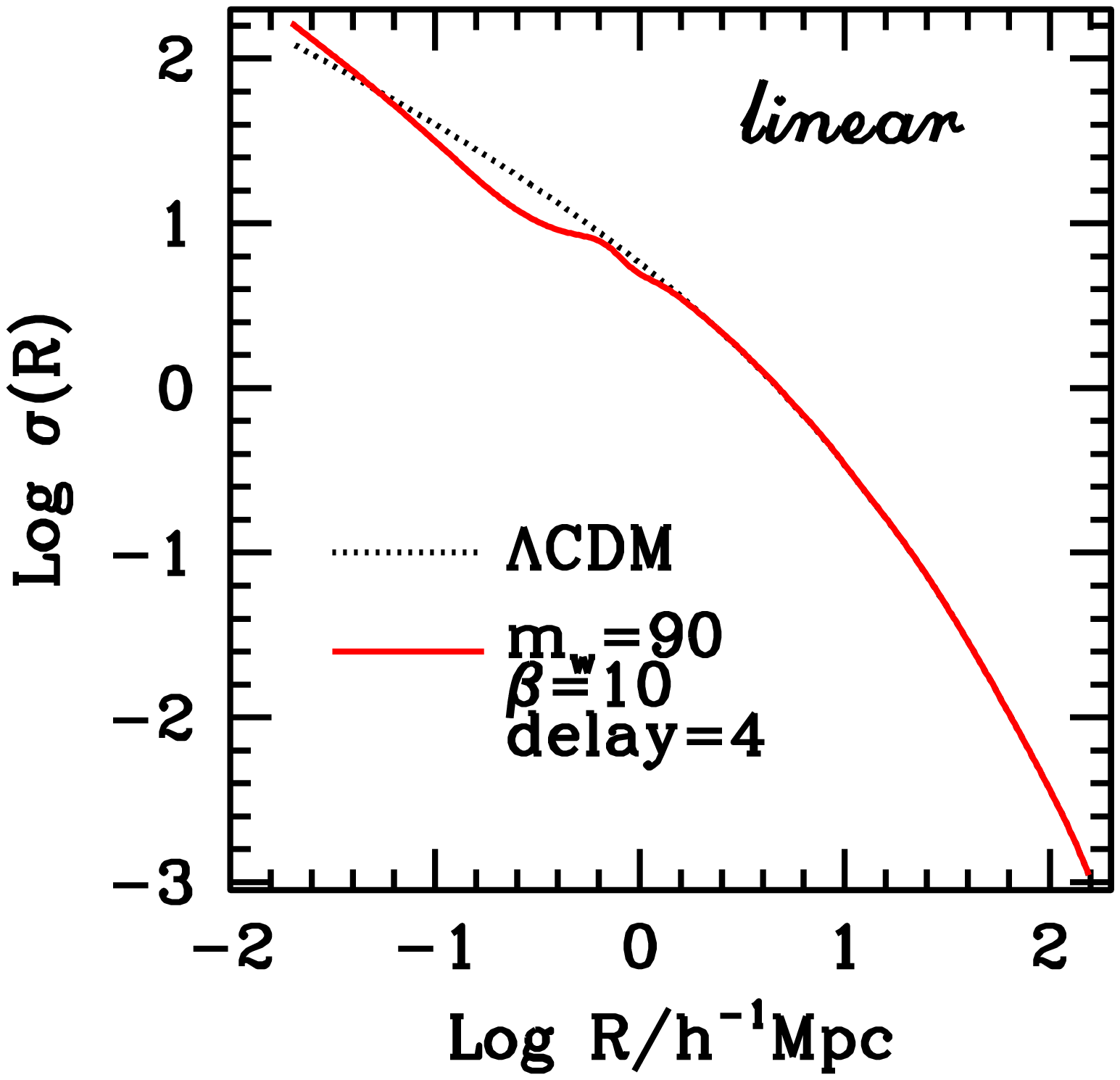}
\end{center}
\caption{{\it Linear} mass variance on the scale $R$ for the model in
  the frame (an exponential window is used). }
\label{ss}
\end{figure}
\section{Fluctuation spectra}\label{sec:Spectra}

The modified version of {\sc cmbfast} enables us to evaluate the
transfer functions and the spectral functions $ \Delta^2(k,z) $ (see
eq.~\ref{d2}).  Figures \ref{tfa} show them at $z=0$, for specific
models, with separate curves for different components.  Results are
compared with $\Lambda$CDM.

For $\log k \lesssim -1$ the spectra of all cosmic components almost
overlap. At greater $k$ values, baryons and WDM exhibit a gap in
respect to $\Lambda$CDM. In the absence of strongly coupled CDM, such
hint to decline would turn into a fast decrease at slightly higher
$k$'s, a characteristic of standard $\Lambda$WDM cosmologies.  Here,
on the contrary, thanks to the action of coupled CDM, we see a fast
recovery. Baryon and WDM spectra therefore re--approach the
$\Lambda$CDM spectrum and, in the case $d=4$, overcome it. In turn,
the (coupled) CDM spectra are constantly above baryon and WDM; for
$d=2$, however, the CDM spectrum does first decline, as baryons and
WDM do, then gradually rises above them; on the contrary, for $d=4$,
the CDM spectrum keeps constantly above $\Lambda$CDM and there is just
a hint of slower increase where baryons and WDM begin their gap.

The most significant point, however, is the comparison with
$\Lambda$CDM. Figures \ref{tfa} show a greater spectral discrepancy
for lower $d$ values. If we however consider $\beta \neq 10$ values
(not shown in the Figures), we have a similar evolution of CDM
fluctuations, for the $\beta$ compensation discussed in Section 4.1,
but equally large CDM fluctuations are less efficient to recreate WDM
and baryon fluctuations, because of the smaller CDM mass attracting
them. Accordingly, the discrepancy with $\Lambda$CDM is even greater,
in spite of CDM having spectra similar to the ones shown here. As a
matter of fact, with $\beta = 10$, the discrepancy of baryon and WDM
spectra with $\Lambda$CDM is not so great, never exceeding 1 order of 
magnitude.

There is another point, already visible in the spectral functions in
Figure \ref{tfa}, but conveniently stressed in Figure \ref{wb}: at high
$k$ the baryon spectrum starts to exceed the WDM spectrum.  It is not
a negligible effect. At $k \simeq 30\, h\, $Mpc$^{-1},$ the baryon
spectrum exceeds average by $\sim 60\, \%$, while tha WDM spectrum
exhibits a deficiency, still in respect to average, by $\sim 25\,
\%$. In Figure \ref{wb} curves referring to models with $d = 2,~3,~4$,
are shown to overlap. The Figure is done for $m_{\rm w} = 90\, $eV; as
expected, a significant increase of $m_{\rm w}$ reduces the gap.

It may be significant to see the shift between SCDEW and $\Lambda$CDM,
when looking at the linear m.s. fluctuation 
\begin{equation}
\sigma_R^2 = \int_0^\infty dk ~\Delta^2(k)\, \,  W^2(kR)
\label{sigma2}
\end{equation}
(mass variance). There are different forms for the window function
$W(kR)$; in this work all computations were coherently performed by
using an exponential window. The $\sigma_R$ behavior is shown in
Figure \ref{ss}, for $d=4$ where, within the models considered here,
it is greatest. As recalled in the frame, this is a {\it linear}
computation, and the slight discrepancies appear on scales $R < 1\,
h^{-1}$Mpc ($\sim 3.5 \times 10^{11}h^{-1} M_\odot$), surely
non--linear today. The expected overall effect, after the extensive
non--linear $k$--mode mixing, is expected to be slightly greater.

\begin{figure}
\begin{center}
\includegraphics[width = 0.47\textwidth]{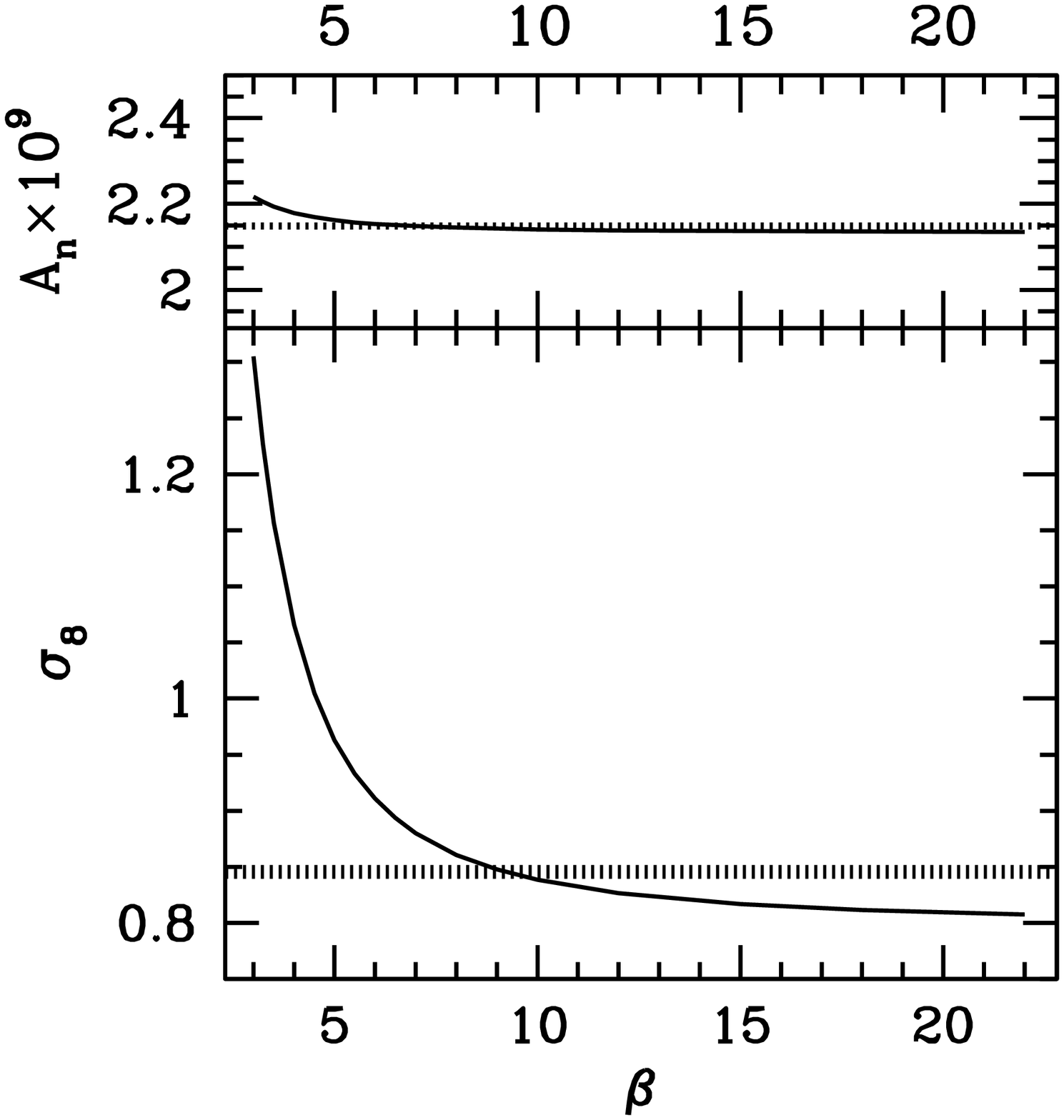}
\end{center}
\vskip -0.5truecm
\caption{Lower plot: $\sigma_8$ dependence on $\beta$ for SC models
  best fitting CMB fluctuations (hence, normalized as shown in the top
  plot). In the SC model selected $m_{\rm w}=90\, $eV. Dashed horizontal
  lines yield normalization and $\sigma_8$ for a $\Lambda$CDM model
  with the same parameters. }
\vskip -.2truecm
\label{sigbet}
\begin{center}
\includegraphics[width = 0.46\textwidth]{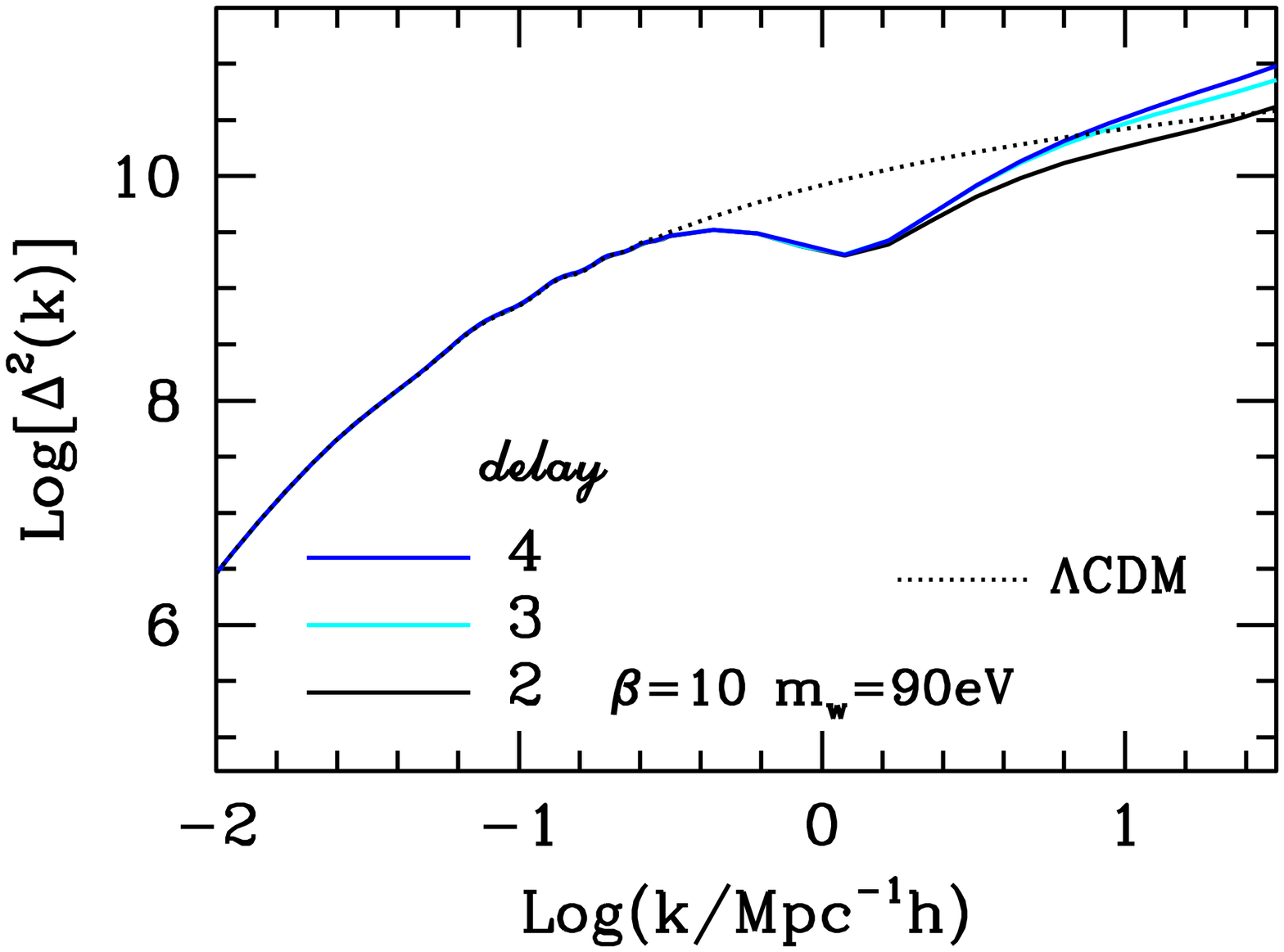}
\end{center}
\vskip -1.truecm
\caption{Dependence of the overall spectrum on the {$d$} parameter.
At $k \simeq 32\, h\, $Mpc$^{-1}$ ($M \simeq 2.8 \times 10^9 h^{-1}M_\odot$)
the ratio between $d=4$ and $d=2$ spectra is $\lesssim 2$ ~.}
\label{sp}
\end{figure}

The expression (\ref{sigma2}) can be also used to study the $\beta$
dependence of $\sigma_8$ (mass variance on the scale of 8$\,
h^{-1}$Mpc) on $\beta$ and $m_{\rm w}$. In Figure \ref{sigbet}, we
keep $m_w$ fixed and study the $\sigma_8$ dependence on $\beta$ for
models with $d=4$. The main finding shown in this plot is that, in
order to approach the $\sigma_8$ value of $\Lambda$CDM, we need $\beta
\simeq 10$.

Let us then outline the difference between total spectra of models
differing for the {$d$} values. As shown in Figure \ref{sp},
discrepancies increase with $k$. For $\log k \simeq 1.5$ ($M \simeq
2.8 \times 10^{9} h^{-1} M_\odot$) the discrepancy is $\sim 2$, when
we compare $d = 4$ and 2~; it becomes $<20\, \%$ when we compare $d$=3
and 4. The above feature however exhibits just a mild $\beta$
dependence (not shown in the Figure): for lower (higher) $\beta $
values, the discrepancy is slightly smaller (greater).

These spectra were accurately considered to choose the SCDEW model
most suitable to study non--linear effects. Coupled CDM
non--linearities are safely milder with $d=2$. (They would be even
milder for $d<2$ but, as shown in Figure 1, the $d$ dependence of the
model becomes stronger and, in particular, a significant gap in
$\rho_\Phi$ behavior becomes unavoidable.)  As discussed in previous
Section, even in the $d=2$ case we expect some impact of forming CDM
non--linearities, probably slightly rising the spectrum, more and more
as $k$ increases. Owing then to the limited difference between
spectra, shown in Figure \ref{sp}, we decided to make use of the $d=4$
spectrum, assuming that, probably, it best represents the $d=2$ case,
once the upward shifts due to forming CDM non--linearities are
included.

\section{Discussion}

A number of experiments are in progress, aiming to inspect the nature
of dark cosmic components, e.g.
BOSS\footnote{http://www.sdss3.org/surveys/boss.php},
HETDEX\footnote{http://hetdex.org/hetdex},
DES\footnote{http://www.darkenergysurvey.org},
LSqST\footnote{http://www.lsst.org} and
Euclid\footnote{www.euclid-ec.org/‎}. Even assuming a linear expression
$w = w_0 + w_1(1-a)$ for DE state parameter, however, errors on $w_0$
and $w_1$ can hardly be pushed below 10 $\, \%$ and some 10$\, \%$,
respectively  \citep[e.g.][]{joachimi}. Better hopes exist to
detect violations of the {\it standard} relation between scale factor
$a$ and growth factor $\cal G$ time dependences.

They could arise from violation of General Relativity if, e.g.,
the gravitational action $R$ (Riemann scalar) is replaced by a suitable
function $f(R)$. In this case the dynamics of any component, baryons
and DM, would be directly modified at large distances. A large deal of
work has deepened this option. N--body simulations have been also
recently performed \citep{puchwein}.

An alternative option is that only DM dynamics is modified, as occurs
in ``ordinary'' coupled DE theories. Also this option has been widely
debated, N-body simulations of these models have been performed since
several years \citep{maccio2004,baldi,LiBarrow,Baldi2012}, finding
that baryon distribution would also change, over very large scales, as
an indirect consequence. Since a few years these models have been
shown to be consistent with data if $\beta \lesssim 0.1~$, limits
being widened if coupling is considered together with a non--vanishing
neutrino mass, even finding a $~\sim 2$--$\sigma$ constraint on the
coupling $\beta$ about $\sim 0.09$
\citep{lavacca,kristiansen,pettorino}.  More recently, the coupling
option was also shown to ease the tension between Planck and Hubble
telescope $H_0$ estimates, allowing \citet{xia} to turn limits into a
$>3$--$ \sigma$ detection, yielding $\beta = 0.078 \pm 0.022$ and,
consistently, $H_0 = 74.8 \pm 2.8$. Also within this option,
significantly affecting the dynamics of the whole DE, future data are
expected to outline apparent $a(t)$--$\cal G$$(t)$ discrepancies.

Strongly Coupled cosmologies (SCDEW models), on the contrary, predict
no change in $a(t)$ or $\cal G$$(t)$ in respect to $\Lambda$CDM and
therefore, {\it a fortiori}, the $a(t)$--$\cal G$$(t)$ relation should
be found do be consistent with $\Lambda$CDM predictions. The very
equation of state of DE remains somehow arbitrary, although state
parameters $w < -1$ would require suitable extentions of the approach
described in this work. All above experiments are therefore expected
to yield results seemingly consistent with $\Lambda$CDM cosmologies.

This point has been accurately verified in this work. In comparison
with BSLV12 and BM14, two further points were outlined here: (i) By
providing a Lagrangian approach to the coupling between $\Phi$ and
$\psi$ fields (DE and CDM, respectively) and stressing the logarithmic
growth of the former one, we outlined the possibility that $\Phi$ is
both inflaton and DE. (ii) The risk of CDM fluctuations reaching an
early non--linear dynamics was also outlined and {\it circumscribed.}

We shall further discuss the (i) point in another work. As far as the
(ii) point is concerned, it is strictly linked to the {\it minimal}
assumption of a constant interaction constant $\beta$.  We tentatively
suggested to overcome such option by admitting $\beta$ to fade, once
it accomplished its aim to allow us an (almost) conformally invariant
cosmic expansion through cosmic ages. Possible mechanisms 
  directly relating conformal invariance break to $\beta$ fading,
however, were not discussed. 
 A number of alternative options are briefly discussed in Appendix
  A.

Altogether, the picture considered remains fully viable and leads to a
phenomenological picture quite close to $\Lambda$CDM. In its
theoretical framework, however, most of the unpleasant fine tunings
and coincidences of $\Lambda$CDM are significantly eased.

SCDEW and $\Lambda$CDM cosmologies are however distinguishable through
a number of observables: First of all, SCDEW suggests a DM particle
with mass $\sim 100\, $eV. If this kind of warm--hot DM replaces the
CDM of $\Lambda$CDM models, in the absence of strong coupling there
would be almost no structure in the Universe. Even the option of mixed
cold and hot--warm DM is far from fitting data on dwarf rotation
curves or large galaxy satellites. Accordingly, {\it discovering a
  sterile neutrino or a gravitino in the 100$\, $eV mass range would
  mark a strong point in favour of SCDEW models.}

The other critical prediction of SCDEW cosmologies is the formation of
early coupled--CDM structures: Coupled--CDM fluctuations $\delta_c$,
to be able to revive baryon and WDM fluctuations, had to be quite
large when WDM particles finally became non--relativistic. This is
what theory predicts and is also the only way to make their density
excess $\delta \rho_c = \rho_c \delta_c$ significant; in fact $\rho_c
\simeq 1/2\beta^2$ or less.  Therefore, $\delta_c$ approaches unity
earlier than other components. If, meanwhile, the CDM--$\Phi$ coupling
fades, baryons and CDM share identical equations of motion and only
their total fluctuation matters. Non--linearity then becomes just a
formal and harmless feature. Should it be not so, bound CDM structures
form. This however occurs over mass scales $\lesssim 10^7
h^{-1}M_\odot$ (let us however recall that, altogether, coupled CDM
density is $\sim 1\, \%$ of the very baryon density, today).

This risks to weaken the predictivity of linear codes. When growing
non linear, CDM fluctuation gravitation becomes stronger than linear
codes compute, speeding up the growth of other components. This effect
is however weakened by $\rho_c$ becoming even smaller than $\sim
1/2\beta^2$, as the Universe turns from radiation to matter dominated,
and by the fact that other component fluctuations have recovered such
a significant amplitude to yield an --at least-- comparable push.
Accordingly, we suggested the extra gravitational push of CDM to bear
effects equivalent to shifting the parameter $d$ from 2 to 4. Although
based on qualitative arguments, this option is not unlikely to
approach the real effect. Accordingly, N--body simulations in the
accociated Paper II are based on the $d=4$ option.
 
All that implies a clear prediction: that early CDM primeval
structures have formed. Although being extremely rare because of the
low overall CDM density ($\sim 1/100$ of baryon density), they could
be the main existing structures at high redshifts ($z \gg 10$) and
might have an effect at later times on the formation of the first
stars and/or cosmic reionization; we plan to address these issues in
future work.
 
In our opinion, however, the strongest argument in favour of SCDEW
models is the apparent easing of fine tunings. 
In particular, a twofold DM is an option often considered on purely
phenomenological bases. However, the {\it true} DM, in SCDEW
cosmologies, is the warm one; CDM is a sort of handyman component,
fist allowing $\Phi$ to keep on its tracker solution, then allowing
inhomogeneities to revive on observational scales, eventually creating
bound systems where it {\it hides} today. The real challenge of these
models is finding out its hideouts. Apart of that, replacing $\Lambda$
with a scalar field might seem not a fresh option.  Also the
independence from I.C., thanks to the presence of a field attractor,
is not new.

In respect to standard quintessence theories, however, SCDEW models
are favoured due to a few strong points: (i) They are apparently independent
from the choice of any specific potential for the $\Phi$ field. (ii)
The quintessential $\Phi$ field is shown to have necessarily been a
significant cosmic component since long time, possibly since the end
of inflation. (iii) Again, the attractor solution it fulfills is not
linked to any potential choice. (iv) The possibility that DE and
inflaton are the same field $\Phi$ can also be pursued: in effect,
since inflation, $\Phi$ had just a logarithmic growth matching, e.g.,
the logaritmic evolution of constants in a Coleman--Weinberg--like
potential, so able to provide potential energy both then and now.

The idea that $\Lambda$CDM is such a successfull model because it {\it
  mimics} an underlying more complex cosmology is not completely
  new. For instance, \cite{ABW} investigated cosmologies
  where DE arises because of the increase of neutrino mass, aiming to
  a scenario substantially indistinguishable from $\Lambda$CDM.
Any such cosmology requires the introduction of extra parameters.
In SCDEW models 2 extra parameters (apart of
the delay $d$) are needed:
the mass of the WDM particle and the coupling $\beta$. However, if we
vary these parameters within a reasonable range, we obtain cosmologies
which could also be {\it mimiced} by sorts of $\Lambda$CDM with a
different choice of its basic parameters. Moreover, the mass $m_{\rm
  w}$ could soon become a non--free parameter, if a particle candidate
is found. In this case, the costraints on $\beta$, deriving from
fluctuation amplitude, CMB spectrum, density parameter choice, etc.,
as illustrated in this paper, are decisive.

\section*{Acknowledgments}

S.A.B. thanks C.I.F.S. (Consorzio Interuniversitario per la Fisica
Spaziale) for its financial support.
 

\vskip 1.8truecm 

\appendix{\bf APPENDIX A: PHYSICAL OPTIONS FOR $\beta$ FADING}

\vglue .6truecm
\noindent
In the CDM--$\Phi$ coupling $C = b/m_p$ one can straightforwardly
introduce a $\Phi$ dependence, yielding $C(\Phi)$; \cite{mainini},
e.g., took $C = b/\Phi$. In the context of SCDEW modes, owing to the
progressive $\Phi$ increase, instead of taking $\beta \propto
(a/a_{dg})^\alpha$, we can assume $\beta \propto
(\Phi/\Phi_{dg})^{\tilde \alpha}$; an expression relating $a_{dg}$ and
$\alpha$ to $\Phi_{dg}$ and $\tilde \alpha$ is then obtainable, so to
achieve similar results.

A smarter option, perhaps, still keeping to field theory, amounts to
replacing the lagrangian (6) in Section~2 by
$$ 
{\cal L}_m = -(\mu f(\Phi/m) + \tilde \mu) \bar \psi \psi~.
\eqno (A1)
$$ 
This modifies the field and $\rho_c$ equations~(5) into
$$
\ddot \Phi + 2{\dot a \over a} \Phi = -a^2 V' +  \rho_c {Ca^2
\over 1+ \exp(C\Phi) \tilde \mu /\mu}
$$ 
$$
\dot \rho_c + 3{\dot a \over a} \rho_c = - \rho_c {C \dot \Phi
\over 1+\exp(C\Phi) \tilde \mu /\mu}
\eqno (A2)
$$ 
where we keep to the expression of $C=b/m_p$ with constant $b$. Until
$b\Phi \ll m_p$ coupling is as usual. But, when $\Phi$ approaches the
Planck scale, the effective coupling is cut off.

A further option, not affecting field theory, can also be envisaged,
although its effectiveness depends on a number of still open
problems. As a matter of fact one can wonder what happens to CDM
nonlinearities after they entry in the horizon. The question is whether the nonlinear
gravitational growth of CDM fluctuations ends up into a relativistic
collapse or into virialized structures. In the former case, one should
not forget that the action of $\Phi$ on CDM can be described by
replacing $G$ with $G^*$, adding the antidrag term, etc., only until a
non relativistic approximation holds. On the contrary, the dynamics of
a relativistic collapse, in the presence of $\Phi$ interactions, is
still an open problem.

We can however formulate two conjectures: (i) The nonlinear
gravitational growth ends up into virialization if the CDM matter
density overcomes a suitable theshold $\bar \rho_{c}$. (ii) If
relativistically collapsed CDM structures form, the $\Phi$ field is
unable to interact with CDM there inside.

Within this context, let us imagine to treat the growth of a spherical
density fluctuation, starting from its entry in the horizon, when it
involves fluctuations of amplitude $\Delta_c \sim \bar \Delta \sim
10^{-5}$ (for CDM and the other cosmic components, respectively). Let
$R = ac$ be its radius, $a$ being the scale factor. At the conformal
initial time $\tau_i$, the radius is $R_i = a_i c_i \sim \tau_i$ with
$\dot c_i = 0$. Let then $x = c/c_i$; at any $\tau > \tau_i$, $x$
shall fulfill the equation
$$
\ddot x = -{1 \over 2} \left[{\Delta_c \gamma \Omega_c \over x^2}+ 2 (1-\Omega_c)
\bar \Delta x \right] {1 \over \tau^2}~,
\eqno (A3)
$$ 
almost exact in the period characterized by self--similar expansion.
The equation is more complex later on, but the key features are the
same. In particular, let us remind that $\Omega_c = 1/2\beta^2$,
$\gamma = 1+4\beta^2/3$.

According to this equation, as expected, the CDM enhancement expands,
enters a non--linear regime, and then starts to recontract, while the
other components still undergo a linear evolution. Recontraction ends
up into a complete gravitational collapse only if the Schwartzshild density
$$
\rho_s \simeq 3 \times 10^{16} {\rm (g/cm^3)} (M/M_\odot)^{-2}
$$ 
is attained before arriving to the critical density $\bar \rho_c$.
Here $M$ is the mass involved in the collapse. Accordingly, at early
times, CDM non linearities can be described as a rippled CDM
distribution. There will however be a time after which CDM
starts to form sorts of Black Holes (BH). 

In the late epochs, therefore, CDM has mostly turned into a suitable
BH distribution, ceasing its interaction with $\Phi$.

\bsp

\label{lastpage}

\end{document}